\documentclass[english]{IEEEtran}
\usepackage[T1]{fontenc}
\usepackage[latin9]{inputenc}
\usepackage{xcolor}
\usepackage{pdfcolmk}
\usepackage{amsthm}
\usepackage{amsmath}
\usepackage{amssymb}
\usepackage{graphicx}
\PassOptionsToPackage{normalem}{ulem}
\usepackage{ulem}

\makeatletter

\providecommand{\tabularnewline}{\\}
\providecolor{lyxadded}{rgb}{0,0,1}
\providecolor{lyxdeleted}{rgb}{1,0,0}

  \theoremstyle{definition}
  \newtheorem{defn}{\protect\definitionname}
  \theoremstyle{remark}
  \newtheorem{rem}{\protect\remarkname}
  \theoremstyle{plain}
  \newtheorem{lem}{\protect\lemmaname}
  \theoremstyle{definition}
  \newtheorem{example}{\protect\examplename}
  \theoremstyle{plain}
  \newtheorem{cor}{\protect\corollaryname}
  \theoremstyle{plain}
  \newtheorem{thm}{\protect\theoremname}

\newtheorem{assumption}{Assumption}
\theoremstyle{definition}
\newtheorem{observation}{Observation}

\@ifundefined{showcaptionsetup}{}{%
 \PassOptionsToPackage{caption=false}{subfig}}
\usepackage{subfig}
\makeatother

\usepackage{babel}
\providecommand{\corollaryname}{Corollary}
\providecommand{\definitionname}{Definition}
\providecommand{\examplename}{Example}
\providecommand{\lemmaname}{Lemma}
\providecommand{\remarkname}{Remark}
\providecommand{\theoremname}{Theorem}

\begin{document}

\title{Hierarchical Radio Resource Optimization for Heterogeneous Networks
with Enhanced Inter-cell Interference Coordination (eICIC)}

\author{{\normalsize An Liu$^{1}$, }\textit{\normalsize Member IEEE}{\normalsize ,
Vincent K. N. Lau$^{1}$,}\textit{\normalsize{} Fellow IEEE}{\normalsize ,
Liangzhong Ruan$^{1}$, Junting Chen$^{1}$, }\textit{\normalsize Member
IEEE}{\normalsize , and Dengkun Xiao$^{2}$\\}$^{1}${\normalsize Department
of Electronic and Computer Engineering, Hong Kong University of Science
and Technology\\$^{2}$Huawei Technologies CO., LTD.}%
\thanks{This work is funded by Huawei Technologies.%
}\vspace{-0.25in}
}
\maketitle
\begin{abstract}
Interference is a major performance bottleneck in Heterogeneous Network
(HetNet) due to its multi-tier topological structure. We propose almost
blank resource block (ABRB) for interference control in HetNet. When
an ABRB is scheduled in a macro BS, a resource block (RB) with blank
payload is transmitted and this eliminates the interference from this
macro BS to the pico BSs. We study a two timescale hierarchical radio
resource management (RRM) scheme for HetNet with \textit{dynamic ABRB}
control. The long term controls, such as dynamic ABRB, are adaptive
to the large scale fading at a RRM server for co-Tier and cross-Tier
interference control. The short term control (user scheduling) is
adaptive to the local channel state information within each BS to
exploit the \textit{multi-user diversity}. The two timescale optimization
problem is challenging due to the exponentially large solution space.
We exploit the sparsity in the \textit{interference graph} of the
HetNet topology and derive structural properties for the optimal ABRB
control. Based on that, we propose a \textit{two timescale alternative
optimization} solution for the user scheduling and ABRB control. The
solution has low complexity and is asymptotically optimal at high
SNR. Simulations show that the proposed solution has significant gain
over various baselines.\end{abstract}

\section{Introduction}

As HetNet provides flexible and efficient topology to boost spectral
efficiency, it has recently aroused immense interest in both academia
and industry. As illustrated in Fig. \ref{fig:Hetnet_2tiers}, a HetNet
consists of a diverse set of regular macro base stations (BS) overlaid
with low power pico BSs. Since this overlaid structure may lead to
severe interference problem, it is extremely critical to control interference
via RRM in HetNet. There has been much research conducted on RRM optimization
for traditional cellular networks. In \cite{Ebrahimi_TIT07_scalecell,Gesbert_TIT11_scalecell},
the authors considered power and user scheduling in single-carrier
cellular networks. In \cite{gesbert2007adaptation,altman2006survey},
the game theoretical approaches are proposed for distributed resource
allocation. In \cite{Stolyar:2008fk}, the authors proposed a dynamic
fractional frequency reuse scheme to combat the inter-sector interference
under a game-based optimization by each sector. The coordinated multipoint
transmission (CoMP) \cite{irmer2011coordinated} is another important
technique to handle the inter-cell interference. For example, in \cite{dahrouj2010coordinated},
the authors exploited the uplink-downlink duality to do joint optimization
of power allocation and beamforming vectors. In \cite{shi2011iteratively},
a WMMSE algorithm is proposed to find a stationary point of the weighted
sum-rate maximization problem for multi-cell downlink systems. While
the above algorithms achieve comparably good performance, they require
global channel state information (CSI) for centralized implementation
\cite{dahrouj2010coordinated} or over-the-air iterations and global
message passing for distributed implementation \cite{shi2011iteratively}.
It is quite controversial whether CoMP is effective or not in LTE
systems due to large signaling overhead, signaling latency, inaccurate
CSIT, and the complexity of the algorithm. 

On the other hand, solutions for traditional cellular networks cannot
be applied directly to HetNet due to the unique difference in HetNet
topology. First, the inter-cell interference in HetNet is more complicated,
e.g., there is \textit{co-tier interference} among the pico BSs and
among the macro BSs as well as the \textit{cross-tier interference}
between the macro and pico BSs. Furthermore, due to load balancing,
some of the mobiles in HetNet may be assigned to a pico BS which is
not the strongest BS \cite{Qualcomm2012_lte} and the mobiles in the
pico cell may suffer from strong interference from the macro BSs.
To solve these problems, some eICIC techniques, such as the ABS control
\cite{Qualcomm2012_lte}, have been proposed in LTE and LTE-A \cite{3gpp_Rel10}.
In \cite{Wang:2012uq}, the authors analyzed the performance for ABS
in HetNet under different cell range extension (RE) biases. However,
they focused on numerical analysis for the existing heuristic eICIC
schemes, which are the baselines of this paper. In \cite{Pang:2012kx},
the authors proposed an algorithm for victim pico user partition and
optimal synchronous ABS rate selection. However, they used a universal
ABS rate for the whole network, and as a result, their scheme could
not adapt to dynamic network loading for different macro cells.

In this paper, we focus on the resource optimization in the downlink
of a HetNet without CoMP%
\footnote{While there are a lot of works using multi-antenna techniques (CoMP)
to mitigate interference in HetNet \cite{dahrouj2010coordinated,Luo_Arxiv12_WMMSE_HetNet},
such approaches require accurate knowledge of at least the cross-link
CSIT at each macro and pico BS, which is not realistic in practice.
As a result, the LTE-A working groups are actively studying eICIC
techniques such as ABS for interference control of HetNet.%
}. We consider \textit{dynamic ABRB} for interference control and dynamic
user scheduling to exploit \textit{multi-user diversity}. The ABRB
is similar to the ABS but it is scheduled over both time and frequency
domain. Unlike \cite{Pang:2012kx}, we do not restrict the ABRB rate
to be the same for all macro BSs and thus a better performance can
be achieved. However, this also causes several new technical challenges
as elaborated below.
\begin{itemize}
\item \textbf{Exponential Complexity for Dynamic ABRB: }Optimization of
ABRB patterns is challenging due to the combinatorial nature and exponentially
large solution space. For example, in a HetNet with $N_{0}$ macro
BSs, there are $2^{N_{0}}$ different ABRB pattern combinations. Hence,
brute force solutions are highly undesirable. 
\item \textbf{Complex Interactions between dynamic user scheduling and dynamic
ABRB}: There is complex coupling between the dynamic user scheduling
and ABRB control. For instance, the ABRB pattern will affect the user
sets eligible for user scheduling. Furthermore, the optimization objective
of ABRB control depends on user scheduling policy and there is no
closed form characterization. 
\item \textbf{Challenges in RRM Architecture:} Most existing solutions for
resource optimization of HetNet requires global knowledge of CSI and
centralized implementations. Yet, such designs are not scalable for
large networks and they are not robust with respect to (w.r.t.) latency
in backhaul. 
\end{itemize}

To address the above challenges, we propose a two timescale control
structure where the long term controls, such as dynamic ABRB, are
adaptive to the large scale fading. On the other hand, the short term
control, such as the user scheduling, is adaptive to the local CSI
within a pico/macro BS. Such a multi-timescale structure allows \textit{Hierarchical
RRM} design, where the long term control decisions can be implemented
on a RRM server for inter-cell interference coordination. The short-term
control decisions can be done locally at each BS with only local CSI.
Such design has the advantages of low signaling overhead, good scalability,
and robustness w.r.t. latency of backhaul signaling. While there are
previous works on two timescale RRM \cite{Wang:2012uq,Pang:2012kx},
those approaches are heuristic (i.e. the RRM algorithms are not coming
from a single optimization problem). Our contribution in this paper
is a formal study of two timescale RRM algorithms for HetNet based
on optimization theory. To overcome the exponential complexity for
ABRB control, we exploit the sparsity in the \textit{interference
graph} of the HetNet topology and derive structural properties for
the optimal ABRB control. Based on that, we propose a \textit{two
timescale alternative optimization} solution for user scheduling and
ABRB control. The algorithm has low complexity and is asymptotically
optimal at high SNR. Simulations show that the proposed solution has
significant performance gain over various baselines.

\textit{Notation}\emph{s}: Let $1\left(\cdot\right)$ denote the indication
function such that $1\left(E\right)=1$ if the event $E$ is true
and $1\left(E\right)=0$ otherwise. For a set $\mathcal{S}$, $\left|\mathcal{S}\right|$
denotes the cardinality of $\mathcal{S}$.

\section{System Model and Hierarchical Resource Control Policies\label{sec:System-Model}}

\subsection{HetNet Topology and Physical Layer Model}

\begin{figure}
\begin{centering}
\includegraphics[width=88mm]{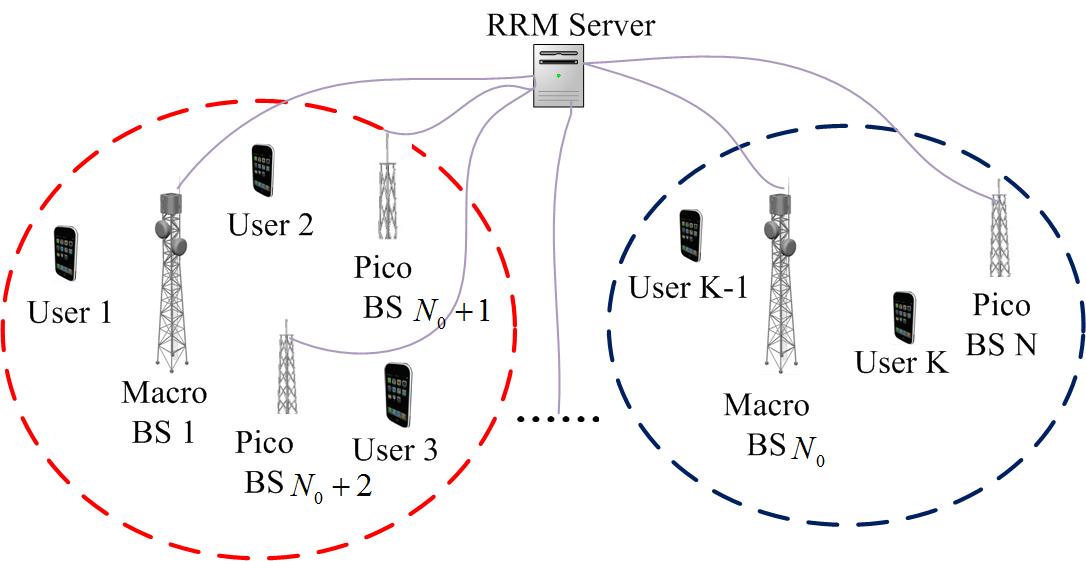}
\par\end{centering}

\caption{\label{fig:Hetnet_2tiers}A two-tier Heterogeneous Network with macro
and pico base stations }
\end{figure}

Consider the downlink of a two-tier HetNet as illustrated in Fig.
\ref{fig:Hetnet_2tiers}. There are $N_{0}$ macro BSs, $N-N_{0}$
pico BSs, and $K$ users, sharing $M$ OFDM subbands. Denote the set
of the macro BSs as $\mathcal{B}_{\textrm{MA}}=\left\{ 1,...,N_{0}\right\} $,
and denote the set of the pico BSs as $\mathcal{B}_{\textrm{PI}}=\left\{ N_{0}+1,...,N\right\} $. 

The HetNet topology (i.e., the network connectivity and CSI of each
link) is represented by a topology graph as defined below.
\begin{defn}
[HetNet Topology Graph]Define the \textit{topology graph} of the
HetNet as a bipartite graph $\mathcal{G}_{T}=\left\{ \mathcal{B},\mathcal{U},\mathcal{E}\right\} $,
where $\mathcal{B}=\left\{ \mathcal{B}_{\textrm{MA}},\mathcal{B}_{\textrm{PI}}\right\} $
denotes the set of all Macro and Pico BS nodes, $\mathcal{U}$ denotes
the set of all user nodes, and $\mathcal{E}$ is the set of all edges
between the BSs and users. An edge $\left(k,n\right)\in\mathcal{E}$
between BS node $n\in\mathcal{B}$ and user node $k\in\mathcal{U}$
represents a wireless link between them. Each edge $\left(k,n\right)\in\mathcal{E}$
is associated with a CSI label $\left\{ h_{m,k,n},\forall m\right\} $,
where $h_{m,k,n}$ represents the channel coefficient between BS $n$
and user $k$ on subband $m$. For each BS node $n$, let $\mathcal{U}_{n}$
denote the set of associated users. For each user node $k$, define
$\mathcal{B}_{k}^{\textrm{MA}}=\left\{ n:n\in\mathcal{B}_{\textrm{MA}},k\notin\mathcal{U}_{n},\left(k,n\right)\in\mathcal{E}\right\} $
as the set of neighbor macro BSs and $\mathcal{B}_{k}^{\textrm{PI}}=\left\{ n:\: n\in\mathcal{B}_{\textrm{PI}},k\notin\mathcal{U}_{n},\left(k,n\right)\in\mathcal{E}\right\} $
as the set of neighbor pico BSs.\hfill \IEEEQED
\end{defn}

\begin{rem}
In the topology graph, $\left(k,n\right)\notin\mathcal{E}$ means
that the path gain between user $k$ and BS $n$ is sufficiently small
compared to the direct link path gain, and thus the interference from
BS $n$ will have negligible effect on the data rate of user $k$.
\end{rem}

We have the following assumption on the channel fading process $\mathbf{H}\left(t\right)=\left\{ h_{m,k,n}\left(t\right)\right\} $.

\begin{assumption}[Two timescale fading model]\label{asm:forY}The
channel fading coefficient has a two timescale structure given by
$h_{m,k,n}\left(t\right)=\sigma_{k,n}\left(t\right)W_{m,k,n}\left(t\right),\:\forall m,k,n$.
The small scale fading process $W_{m,k,n}\left(t\right)$ is identically
distributed w.r.t. the subframe and subband indices ($t,m$), and
it is i.i.d. w.r.t. user and BS indices ($k,n$). Moreover, for given
$t,m,k,n$, $W_{m,k,n}\left(t\right)$ is a continuous random variable.
The large scale fading process $\sigma_{k,n}\left(t\right)>0$ is
assumed to be a slow ergodic process (i.e., $\sigma_{k,n}\left(t\right)$
remains constant for several \textit{super-frames}%
\footnote{One super-frame consists of $L_{S}$ subframes%
}.) according to a general distribution.

\end{assumption}

The two timescale fading model has been adopted in many standard channel
models. The large scale fading $\sigma_{k,n}\left(t\right)$ is usually
caused by path loss and shadow fading, which changes much slowly compared
to the small scale fading.

We consider the following \textit{biased cell selection} mechanism
to balance the loading between macro and pico BSs \cite{Qualcomm2012_lte}.
Let $b_{k}$ denote the serving BS of user $k$. Let $\beta>1$ denote
the cell selection bias and let $P_{n}$ denote the transmit power
of BS $n$ on a single sub-band. Let $\tilde{n}_{m}=\underset{1\leq n\leq N_{0}}{\textrm{argmax}}\: P_{n}\sigma_{k,n}^{2}$
and $\tilde{n}_{p}=\underset{N_{0}+1\leq n\leq N}{\textrm{argmax}}\: P_{n}\sigma_{k,n}^{2}$
respectively be the strongest macro BS and pico BS for user $k$.
If $\beta P_{\tilde{n}_{p}}\sigma_{k,\tilde{n}_{p}}^{2}\geq P_{\tilde{n}_{m}}\sigma_{k,\tilde{n}_{m}}^{2}$,
user $k$ will be associated to pico cell $\tilde{n}_{p}$, i.e.,
$b_{k}=\tilde{n}_{p}$; otherwise $b_{k}=\tilde{n}_{m}$.

If a user only has a single edge with its serving BS, it will not
receive inter-cell interference from other BSs and thus its performance
is noise limited; otherwise, it will suffer from strong inter-cell
interference if any of its neighbor BSs is transmitting data and thus
its performance is interference limited. This insight is useful in
the control algorithm design later and it is convenient to formally
define the interference and noise limited users.
\begin{defn}
[Interference/Noise Limited User]\label{def:INuser}If a user $k$
has a single edge with its serving BS $b_{k}$ only, i.e., $\sum_{n=1}^{N}1\left(\left(k,n\right)\in\mathcal{E}\right)=1$,
then it is called a \textit{noise limited user} (N-user); otherwise,
it is called an \textit{interference limited user} (I-user).
\end{defn}

Fig. \ref{fig:ABS_intro} illustrates an example of the HetNet topology
graph. In Fig. \ref{fig:ABS_intro}(a), an arrow from a BS to a user
indicates a direct link and the dash circle indicates the coverage
area of each BS. An I-user which lies in the coverage area of a macro
BS is connected to this macro BS, while a N-user does not have connections
with the neighbor macro BSs in the topology graph as illustrated in
Fig. \ref{fig:ABS_intro}(b).

\subsection{Two Timescale Hierarchical Radio Resource Control Variables}

We consider a two timescale hierarchical RRM control structure where
the control variables are partitioned into \textit{long-term} and
\textit{short-term} control variables. The long-term control variables
are adaptive to the large scale fading $\mathbf{\Sigma}$ and they
are implemented at the Radio Resource Management Server (RRMS). The
short-term control variables are adaptive to the instantaneous CSI
$\mathbf{H}$ and they are implemented locally at each macro/pico
BS.

\subsubsection{Dynamic ABRB Control for Interference Coordination (Long-term control)}

\begin{figure}
\begin{centering}
\subfloat[Physical HetNet Topology and ABRB Transmissions.]{\centering{}\includegraphics[width=88mm]{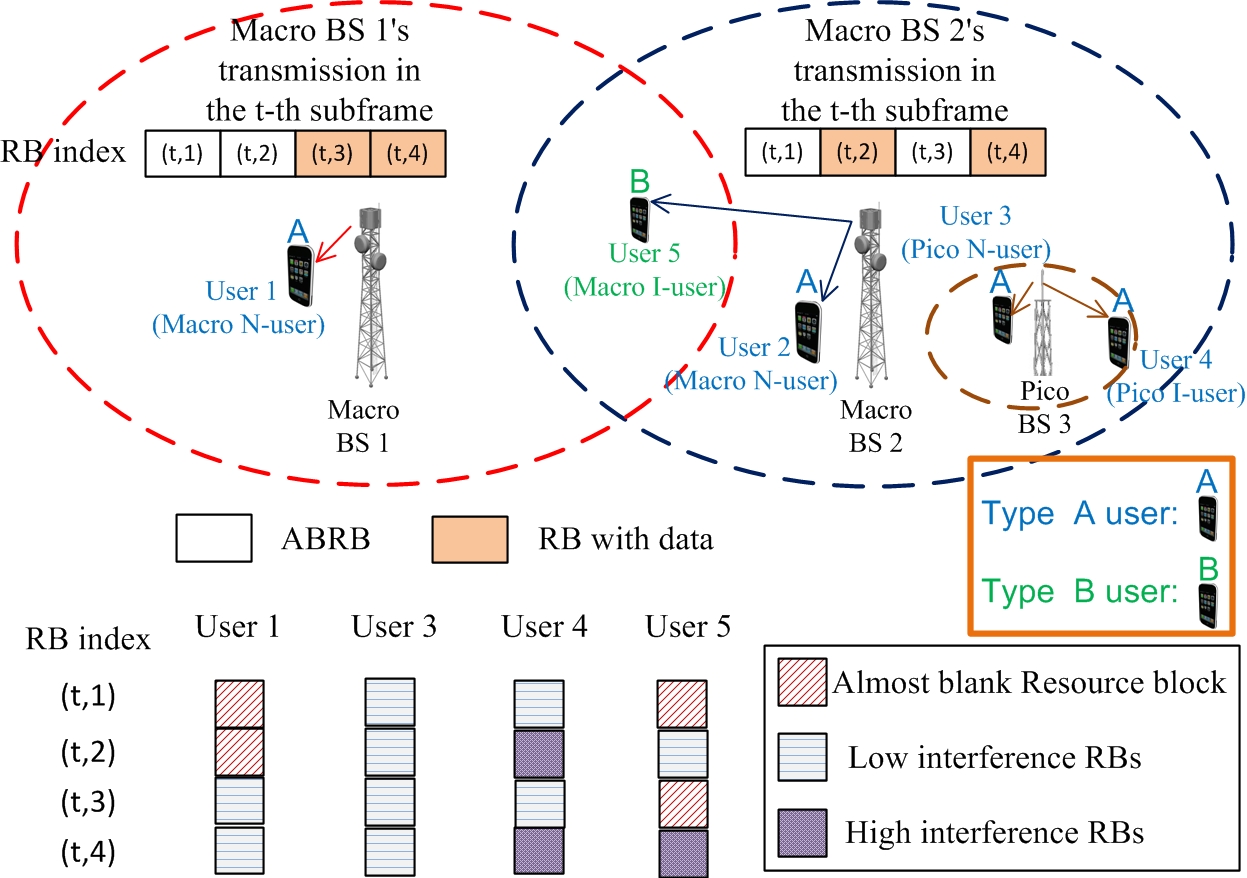}}
\par\end{centering}

\begin{centering}
\subfloat[The corresponding HetNet Topology Graph $\mathcal{G}_{T}=\left\{ \mathcal{B},\mathcal{U},\mathcal{E}\right\} $,
where $\mathcal{B}=\left\{ 1,2,3\right\} $, $\mathcal{U}=\left\{ 1,2,3,4,5\right\} $
and $\mathcal{E}=\left\{ \left(1,1\right),\left(1,5\right),\left(2,2\right),\left(2,4\right),(2,5),\left(3,3\right),\left(3,4\right)\right\} $. ]{\centering{}\includegraphics[width=88mm]{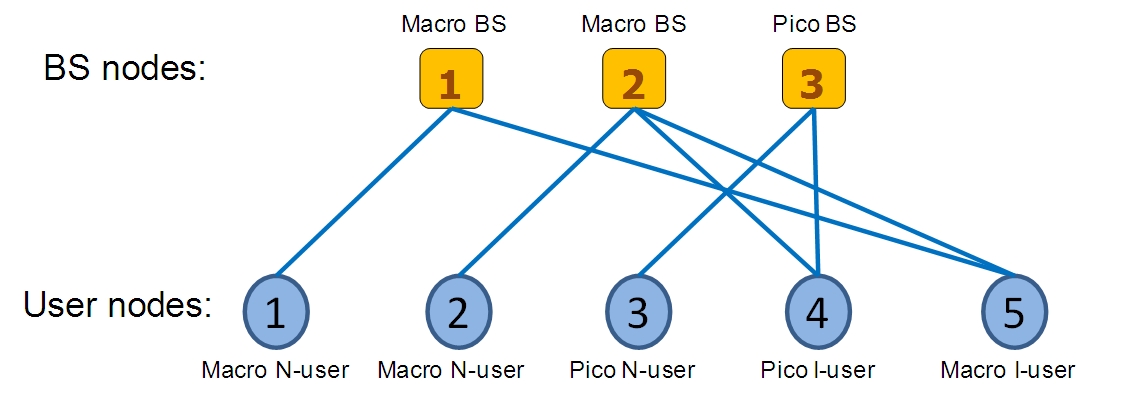}}
\par\end{centering}

\caption{\label{fig:ABS_intro}{\small An example of topology graph and interference
coordination using ABRB. We focus on the $t$-th subframe. The ABRB
can be used to control the interference from macro BSs to pico cell
users. For example, in the $\left(t,1\right)$-th RB (i.e., subframe
$t$ and subband $1$) and $(t,3)$-th RB, the neighbor macro BS 2
of the pico BS 3 transmits ABRB and the pico cell I-user $4$ perceives
a low interference RB. The ABRB can also be used to control the interference
between macro cell users. For example, in the $\left(t,2\right)$-th
RB, macro BS 1 transmits ABRB and the macro cell I-user 5 perceives
a low interference RB.}}
\end{figure}

ABS is introduced in LTE systems \cite{Qualcomm2012_lte} for interference
mitigation among control channels in HetNet. It can also be used to
control the \textit{co-Tier} and \textit{cross-tier interference}
among the data channels. In LTE systems, ABS is only scheduled over
time domain. In this paper, we consider dynamic ABRB control for interference
coordination. The ABRB is similar to ABS but it is scheduled over
both time and frequency domain. It is a generalization of ABS and
enables more fine-grained resource allocation. When an ABRB is scheduled
in a macro BS, a RB with blank payload will be transmitted at a given
frequency and time slice and this eliminates the interference from
this macro BS to the pico BSs and the adjacent macro BSs. Hence, as
illustrated in Fig. \ref{fig:ABS_intro}, scheduling ABRB over both
time and frequency domain allows us to control both the \textit{macro-macro}
BS and \textit{macro-pico} BS interference. We want to control the
ABRB dynamically w.r.t. the large scale fading because the optimal
ABRB pattern depends on the HetNet topology graph. For example, when
there are a lot of pico cell I-users, we should allocate more ABRBs
at the macro BS to support more pico cell I-users. On the other hand,
when there are only a few pico cell I-users, we should allocate less
ABRBs to improve the spatial spectrum efficiency.

For any given subframe, define $a_{m,n}\in\left\{ 0,1\right\} $ to
indicate if ABRB is scheduled ($a_{m,n}=0$) for subband $m$ at macro
BS $n$. Let $\mathbf{a}_{m}=\left[a_{m,1},...,a_{m,N_{0}}\right]^{T}\in\mathcal{A}$
be the ABRB pattern vector for subband $m$ and $\mathcal{A}$ is
the set of all possible ABRB patterns%
\footnote{Since each of the $N_{0}$ macro BSs can either schedule an ABRB or
not for the $\left(t,m\right)$-th RB (i.e., subframe $t$ and subband
$m$), there are $2^{N_{0}}$ possible ABRB patterns. Hence, the size
of $\mathcal{A}$ is $2^{N_{0}}$.%
}. In the proposed dynamic ABRB control, each macro BS is allowed to
dynamically change the ratio of ABRB transmission on each subband
and this ratio can be any positive real number. To facilitate implementation,
we consider randomized ABRB control policy as defined below.
\begin{defn}
[Randomized ABRB Control Policy] An ABRB control policy of the $m$-th
subband $Q_{m}$ is a mapping from the ABRB pattern space $\mathcal{A}$
to a probability in {[}0,1{]}. At any subframe, the instantaneous
ABRB pattern vector for subband $m$ is stochastically determined
according to the probabilities $Q_{m}\left(\mathbf{a}_{m}\right),\:\forall\mathbf{a}_{m}\in\mathcal{A}$,
where $Q_{m}\left(\mathbf{a}_{m}\right)$ denote the probability that
the $m^{\textrm{th}}$ subband is in ABRB pattern $\mathbf{a}_{m}$. 
\end{defn}

\subsubsection{Subband Partitioning Control for Structural ABRB Design (Long-term
control)}

To facilitate structural ABRB design, we partition the users into
two types.
\begin{defn}
[Partitioning of User Set]\label{def:user-set-partition} The mobile
user set is partitioned into two subsets $\mathcal{U}=\left\{ \mathcal{U}_{A},\mathcal{U}_{B}\right\} $,
where $\mathcal{U}_{A}$ denotes the set of \textit{Type A users}
and is defined as
\[
\mathcal{U}_{A}=\left\{ k:\: b_{k}\in\mathcal{B}_{\textrm{PI}}\right\} \cup\left\{ k:\:\sum_{n=1}^{N}1\left(\left(k,n\right)\in\mathcal{E}\right)=1\right\} ,
\]
and $\mathcal{U}_{B}=\mathcal{U}/\mathcal{U}_{A}$ denotes the set
of \textit{Type B users}.\hfill \IEEEQED
\end{defn}

The Type A users include all pico cell users and macro cell N-users,
while the Type B users include all macro cell I-users. For Type A
users, it will not lose optimality by imposing a \textit{synchronous
ABRB structure} where the transmissions of the ABRB at all macro BSs
are aligned as much as possible. The formal definition of the synchronous
ABRB structure is given in Theorem \ref{thm:ABSreduce}. As will be
shown in Theorem \ref{thm:ABSreduce}, if there is only Type A users,
imposing the synchronous ABRB structure can dramatically reduce the
number of ABRB control variables from exponential large ($2^{N_{0}}$)
to only $N_{0}$ and this complex reduction is achieved without loss
of optimality. On contrast, the performance of the macro cell I-users
is very poor under the synchronous ABRB structure because aligning
the data transmissions of all macro BSs will cause strong inter-cell
interference for macro cell I-users. Motivated by these observations,
we partition the $M$ subbands into two groups, namely $\mathcal{M}_{A}$
and $\mathcal{M}_{B}$, and use different ABRB control policies for
type A and type B users on these two groups of subbands respectively.
The variable $q_{s}=\left|\mathcal{M}_{A}\right|/M$ controls the
fraction of \textit{Type A} subbands.

\subsubsection{Dynamic User Scheduling for Multi-user Diversity (Short-term control)}

At each subframe, each BS $n$ dynamically selects a user from $\mathcal{U}_{n}$
for each subband $m$ based on the knowledge of current ABRB pattern
$\mathbf{a}_{m}$ and channel realization $\mathbf{H}_{m}$ to exploit
\textit{multi-user diversity}. Let $\rho_{m,k}\in\left\{ 0,1\right\} $
be the user scheduling variable (of user $k$ at BS $b_{k}$) of subband
$m$ and $\rho_{m}^{n}=\left[\rho_{m,k},k\in\mathcal{U}_{n}\right]$
be the associated vectorized variable. The set of all feasible user
scheduling vectors at BS $n$ for the $m$-th subbands with ABRB pattern
$\mathbf{a}_{m}$ is given by{\small 
\begin{eqnarray*}
\Gamma_{m}^{n}\left(\mathbf{a}_{m}\right)=\;\;\;\;\;\;\;\;\;\;\;\;\;\;\;\;\;\;\;\;\;\;\;\;\;\;\;\;\;\;\;\;\;\;\;\;\;\;\;\;\;\;\;\;\;\;\;\;\;\;\;\;\;\;\;\;\;\;\;\;\;\;\;\;\;\;\\
\begin{cases}
\left\{ \rho_{m}^{n}:\sum_{k\in\mathcal{U}_{n}}\mathbf{\rho}_{m,k}\leq1;\mathbf{\rho}_{m,k}=0,\textrm{if}\: k\in\mathcal{U}_{A}^{0}\left(\mathbf{a}_{m}\right)\right\} ,m\in\mathcal{M}_{A}\\
\left\{ \rho_{m}^{n}:\sum_{k\in\mathcal{U}_{n}}\mathbf{\rho}_{m,k}\leq1;\mathbf{\rho}_{m,k}=0,\textrm{if}\: k\in\mathcal{U}_{B}^{0}\left(\mathbf{a}_{m}\right)\right\} ,m\in\mathcal{M}_{B}
\end{cases}
\end{eqnarray*}
}where $\mathcal{U}_{A}^{0}\left(\mathbf{a}_{m}\right)=\mathcal{U}_{B}\cup\left\{ k:\: b_{k}\in\mathcal{B}_{\textrm{MA}}\:\textrm{and}\: a_{m,b_{k}}=0\right\} \cup\left\{ k:\: b_{k}\in\mathcal{B}_{\textrm{PI}};\:\textrm{and}\:\sum_{n\in\mathcal{B}_{k}^{\textrm{MA}}}a_{m,n}>0\right\} $
is the set of users that cannot be scheduled on a Type A subband under
ABRB pattern $\mathbf{a}_{m}$; and $\mathcal{U}_{B}^{0}\left(\mathbf{a}_{m}\right)=\mathcal{U}_{A}\cup\left\{ k:\: b_{k}\in\mathcal{B}_{\textrm{PI}};\:\textrm{and}\: a_{m,b_{k}}=0\right\} $.
The physical meaning of $\Gamma_{m}^{n}\left(\mathbf{a}_{m}\right)$
is elaborated below. First, if a macro BS is transmitting ABRB, none
of its associated users can be scheduled for transmission. Moreover,
due to large cross-tier interference from macro BSs, a pico cell I-user
cannot be scheduled for transmission if any of its neighbor macro
BSs $n\in\mathcal{B}_{k}^{\textrm{MA}}$ is transmitting data subframe
(i.e., $\sum_{n\in\mathcal{B}_{k}^{\textrm{MA}}}a_{m,n}>0$). As will
be seen in Section \ref{sub:Structural-Properties-ofPA}, explicitly
imposing this user scheduling constraint for the pico cell I-users
is useful for the structural ABRB design.

The user scheduling policy $\pi_{m}$ of the $m$-th sub-band is defined
below.
\begin{defn}
[User Scheduling Policy]\label{def:Randomized-User-Scheduling}
A user scheduling policy of the $n$-th BS and $m$-th sub-band $\pi_{m}^{n}$
is a mapping : $\mathcal{A}\times\mathcal{H}\longrightarrow\Gamma_{m}^{n}\left(\mathbf{a}_{m}\right)$,
where $\mathcal{H}$ is the CSI space. Specifically, under the ABRB
pattern $\mathbf{a}_{m}$ and CSI realization $\mathbf{H}_{m}$, the
user scheduling vector of BS $n$ is given by $\rho_{m}^{n}=\pi_{m}^{n}\left(\mathbf{a}_{m},\mathbf{H}_{m}\right)$.
Let $\pi_{m}=\left\{ \pi_{m}^{n},n=1,...,N\right\} $ denote the overall
user scheduling policy on sub-band $m$.\hfill \IEEEQED
\end{defn}

\section{Two Timescale Hierarchical RRM Design\label{sec:Two-Timescale-RRM}}

\subsection{RRM Optimization Formulation}

Assuming perfect CSI at the receiver (CSIR) and treating interference
as noise, the instantaneous data rate of user $k$ is given by:
\begin{equation}
r_{k}\left(\left\{ \mathbf{a}_{m}\right\} ,\mathbf{H},\left\{ \rho_{m}^{b_{k}}\right\} \right)=\sum_{m\in\mathcal{M}\left(k\right)}\mathcal{I}_{m,k}\left(\mathbf{a}_{m},\mathbf{H}_{m},\rho_{m}^{b_{k}}\right),\label{eq:rateko}
\end{equation}
where $\mathcal{M}\left(k\right)=\mathcal{M}_{A},\forall k\in\mathcal{U}_{A}$,
$\mathcal{M}\left(k\right)=\mathcal{M}_{B},\forall k\in\mathcal{U}_{B}$;
$\mathcal{I}_{m,k}\left(\mathbf{a}_{m},\mathbf{H}_{m},\rho_{m}^{b_{k}}\right)=\rho_{m,k}\log\left(1+\frac{\left|h_{m,k,b_{k}}\right|^{2}P_{b_{k}}}{\Omega_{m,k}}\right)$
is the mutual information of user $k$ contributed by the $m$-th
subband; and $\Omega_{m,k}=1+\sum_{n\in\mathcal{B}_{k}^{\textrm{MA}}}a_{m,n}\left|h_{m,k,n}\right|^{2}P_{n}+1\left(m\in\mathcal{M}_{A}\right)\sum_{n\in\mathcal{B}_{k}^{\textrm{PI}}}\left|h_{m,k,n}\right|^{2}P_{n}$
is the interference-plus-noise power at user $k$ on subband $m$.

For a given policy $\Lambda=\left\{ q_{s},\left\{ Q_{m}\right\} ,\left\{ \pi_{m}\right\} \right\} $
and large scale fading state $\mathbf{\Sigma}\triangleq\left\{ \sigma_{k,n}\right\} $,
the average data rate of user $k$ is given by:

\[
\overline{r}_{k}\left(\Lambda\right)=\textrm{E}\left[\left.r_{k}\left(\left\{ \mathbf{a}_{m}\right\} ,\mathbf{H},\left\{ \rho_{m}\right\} \right)\right|\mathbf{\Sigma}\right]=\sum_{m\in\mathcal{M}\left(k\right)}\overline{\mathcal{I}}_{m,k}\left(Q_{m},\pi_{m}\right),
\]
where the average mutual information on subband $m$ is 
\begin{equation}
\overline{\mathcal{I}}_{m,k}\left(Q_{m},\pi_{m}\right)=\sum_{\mathbf{a}\in\mathcal{A}}Q_{m}\left(\mathbf{a}\right)I_{m,k}\left(\pi_{m},\mathbf{a}\right),\label{eq:IbA}
\end{equation}
and $I_{m,k}\left(\pi_{m},\mathbf{a}\right)=\textrm{E}\left[\left.\mathcal{I}_{m,k}\left(\mathbf{a},\mathbf{H}_{m},\pi_{m}^{b_{k}}\left(\mathbf{a},\mathbf{H}_{m}\right)\right)\right|\mathbf{\Sigma},\mathbf{a}\right]$.
For conciseness, the ABRB pattern $\mathbf{a}_{m}$ for a specific
subband $m$ is denoted as $\mathbf{a}=\left[a_{1},...,a_{N_{0}}\right]$
when there is no ambiguity.

The performance of the HetNet is characterized by a utility function
$U\left(\overline{\mathbf{r}}\right)$, where $\overline{\mathbf{r}}=\left[\overline{r}_{1},...,\overline{r}_{K}\right]$
is the average rate vector. We make the following assumptions on $U\left(\overline{\mathbf{r}}\right)$.

\begin{assumption}[Assumptions on Utility]\label{asm:Ufun}The utility
function can be expressed as $U\left(\overline{\mathbf{r}}\right)\triangleq\sum_{k=1}^{K}w_{k}u\left(\overline{r}_{k}\right)$,
where $w_{k}\geq0$ is the weight for user $k$, $u\left(\cdot\right)$
is assumed to be a concave and increasing function. Moreover, for
any $c,r\geq0$ such that $cr$ and $r$ belongs to the domain of
$u\left(\cdot\right)$, $u\left(r\right)$ satisfies
\[
u\left(cr\right)=f\left(c\right)u\left(r\right)+g(c),
\]
where $f\left(c\right)>0$ and $g(c)$ are some scalar functions of
$c$.

\end{assumption}

The above assumption is imposed to facilitate the problem decomposition
in Section \ref{sub:Problem-Decomposition}. This utility function
captures a lot of interesting cases below.
\begin{itemize}
\item \textbf{Weighted Sum Throughput:} The utility function is $U\left(\overline{\mathbf{r}}\right)=\sum_{k=1}^{K}w_{k}\overline{r}_{k}$.
\item \textbf{$\alpha$-Fair \cite{Mo_TACM00_alfafair}: }$\alpha$-Fair
can be used to compromise between the fairness to users and the utilization
of resources. The utility function is\textbf{ }
\begin{equation}
U\left(\overline{\mathbf{r}}\right)=\begin{cases}
\sum_{k=1}^{K}\textrm{log}\left(\overline{r}_{k}\right), & \alpha=1,\\
\sum_{k=1}^{K}\left(1-\alpha\right)^{-1}\overline{r}_{k}^{1-\alpha}, & \textrm{otherwise}.
\end{cases}\label{eq:alffair}
\end{equation}

\item \textbf{Proportional Fair \cite{Kelly_OPR98_PFS}:} This is a special
case of $\alpha$-Fair when $\alpha=1$.
\end{itemize}

Due to the statistical symmetry of the $M$ subbands, there is no
loss of optimality to consider symmetric policy $\Lambda^{s}=\left\{ q_{s},Q_{A},Q_{B},\pi_{A},\pi_{B}\right\} $,
where $Q_{m}=Q_{A}$ ($Q_{B}$) and $\pi_{m}=\pi_{A}$ ($\pi_{B}$)
if $m\in\mathcal{M}_{A}$ ($\mathcal{M}_{B}$). 
\begin{lem}
[Optimality of Symmetric Policy]\label{lem:Optimality-of-Symmetric}There
exists a symmetric policy $\Lambda^{s*}=\left\{ q_{s}^{*},Q_{A}^{*},Q_{B}^{*},\pi_{A}^{*},\pi_{B}^{*}\right\} $
such that it is the optimal solution of the following optimization
problem:
\end{lem}
\begin{eqnarray}
\widetilde{\mathcal{P}}\left(\mathcal{G}_{T}\right): & \max_{\Lambda}\: U\left(\overline{\mathbf{r}}\left(\Lambda\right)\right)\nonumber \\
\textrm{s.t.} & \sum_{\mathbf{a}\in\mathcal{A}}Q_{m}\left(\mathbf{a}\right)=1;\: Q_{m}\left(\mathbf{a}\right)\geq0, & \forall m.\label{eq:QecCon1}
\end{eqnarray}

Please refer to Appendix \ref{Proof-of-Lemma symP} for the proof.

Moreover, we have $\overline{\mathcal{I}}_{m,k}\left(Q_{A},\pi_{A}\right)=\overline{\mathcal{I}}_{m^{'},k}\left(Q_{A},\pi_{A}\right),\:\forall m,m^{'}\in\mathcal{M}_{A}$
and $\overline{\mathcal{I}}_{m,k}\left(Q_{B},\pi_{B}\right)=\overline{\mathcal{I}}_{m^{'},k}\left(Q_{B},\pi_{B}\right),\:\forall m,m^{'}\in\mathcal{M}_{B}$.
As a result, the utility function under a symmetric policy $\Lambda^{s}$
can be expressed as:
\begin{eqnarray*}
U\left(\Lambda^{s}\right) & = & f\left(Mq_{s}\right)U_{A}\left(Q_{A},\pi_{A}\right)+g\left(M\left(1-q_{s}\right)\right)\sum_{k\in\mathcal{U}_{B}}w_{k}\\
 & + & f\left(M\left(1-q_{s}\right)\right)U_{B}\left(Q_{B},\pi_{B}\right)+g\left(Mq_{s}\right)\sum_{k\in\mathcal{U}_{A}}w_{k},
\end{eqnarray*}
where $U_{A}\left(Q_{A},\pi_{A}\right)=\sum_{k\in\mathcal{U}_{A}}w_{k}u\left(\overline{\mathcal{I}}_{m_{A},k}\left(Q_{A},\pi_{A}\right)\right)$,
$U_{B}\left(Q_{B},\pi_{B}\right)=\sum_{k\in\mathcal{U}_{B}}w_{k}u\left(\overline{\mathcal{I}}_{m_{B},k}\left(Q_{B},\pi_{B}\right)\right)$,
and $m_{A}\in\mathcal{M}_{A}$ ($m_{B}\in\mathcal{M}_{B}$) can be
any Type A (Type B) subband. Finally, for a given HetNet topology
graph $\mathcal{G}_{T}=\left\{ \mathcal{B},\mathcal{U},\mathcal{E}\right\} $,
the two timescale RRM optimization is given by:

\begin{eqnarray}
\mathcal{P}\left(\mathcal{G}_{T}\right): & \max_{\Lambda^{s}}\: U\left(\Lambda^{s}\right)\nonumber \\
\textrm{s.t.} & \sum_{\mathbf{a}\in\mathcal{A}}Q_{A}\left(\mathbf{a}\right)=1;\: Q_{A}\left(\mathbf{a}\right)\geq0, & \forall\mathbf{a}\in\mathcal{A}\label{eq:QecCon1A}\\
 & \sum_{\mathbf{a}\in\mathcal{A}}Q_{B}\left(\mathbf{a}\right)=1;\: Q_{B}\left(\mathbf{a}\right)\geq0, & \forall\mathbf{a}\in\mathcal{A}\label{eq:QecCon2A}
\end{eqnarray}
where (\ref{eq:QecCon1A}) and (\ref{eq:QecCon2A}) ensure that $Q_{A}\left(\cdot\right)$
and $Q_{B}\left(\cdot\right)$ satisfy the definition of probability
mass function (pmf).

\subsection{Problem Decomposition\label{sub:Problem-Decomposition}}

Using primal decomposition, problem $\mathcal{P}\left(\mathcal{G}_{T}\right)$
can be decomposed into the following subproblems.

\textbf{Subproblem A }(\textbf{\small Cross-Tier Interference Control}):
Optimization of ABRB $Q_{A}$ and user scheduling $\pi_{A}$. 
\[
\mathcal{P}_{A}\left(\mathcal{G}_{T}\right):\: U_{A}^{*}\triangleq\max_{Q_{A},\pi_{A}}U_{A}\left(Q_{A},\pi_{A}\right),\:\textrm{s.t.}\:(\ref{eq:QecCon1A})\:\textrm{holds}.
\]

\textbf{Subproblem B }(\textbf{\small Co-Tier Interference Control}):\textbf{
}Optimization of ABRB $Q_{B}$ and user scheduling $\pi_{B}$.
\[
\mathcal{P}_{B}\left(\mathcal{G}_{T}\right):\: U_{B}^{*}\triangleq\max_{Q_{B},\pi_{B}}U_{B}\left(Q_{B},\pi_{B}\right),\:\textrm{s.t.}\:(\ref{eq:QecCon2A})\:\textrm{holds}.
\]

\textbf{Subproblem C} (\textbf{\small Subband Partitioning}): Optimization
of subband partitioning $q_{s}$.

\begin{eqnarray*}
\mathcal{P}_{C}\left(\mathcal{G}_{T}\right): & \max_{q_{s}} & f\left(Mq_{s}\right)U_{A}^{*}+g\left(M\left(1-q_{s}\right)\right)\sum_{k\in\mathcal{U}_{B}}w_{k}\\
 &  & +f\left(M\left(1-q_{s}\right)\right)U_{B}^{*}+g\left(Mq_{s}\right)\sum_{k\in\mathcal{U}_{A}}w_{k}.
\end{eqnarray*}
Note that the solution of $\mathcal{P}_{A}$/$\mathcal{P}_{B}$ is
independent of the value of $q_{s}$ because both $U_{A}\left(Q_{A},\pi_{A}\right)$
and $U_{B}\left(Q_{B},\pi_{B}\right)$ are independent of $q_{s}$.
After solving $\mathcal{P}_{A}$ and $\mathcal{P}_{B}$, the optimal
$q_{s}$ can be easily solved by bisection search. On the other hand,
the optimization of $\pi_{A}$/$\pi_{B}$ is a stochastic optimization
problem because the $\overline{\mathcal{I}}_{m_{A},k}$/$\overline{\mathcal{I}}_{m_{B},k}$
involves stochastic expectation over CSI realizations and they do
not have closed form characterization. Furthermore, the number of
ABRB control variables in $\mathcal{P}_{A}$/$\mathcal{P}_{B}$ is
exponential w.r.t. the number of macro BSs $N_{0}$. We shall tackle
these challenges in Section \ref{sec:Solution-of-PA} and \ref{sub:Solution-of-SubproblemPB}.

\section{Two Timescale Hierarchical Solution for Cross-Tier Interference Control
(Subproblem A)\label{sec:Solution-of-PA}}

In this section, we first derive structural properties of $\mathcal{P}_{A}$
and reformulate $\mathcal{P}_{A}$ into a simpler form with reduced
solution space. Then, we develop an efficient algorithm for $\mathcal{P}_{A}$.

We require the following assumption to derive the results in this
section.

\begin{assumption}[Assumptions for pico cell I-users]\label{asm:picoIedge}For
any $n\in\mathcal{B}_{\textrm{PI}}$, let $\mathcal{U}_{n}^{I}=\left\{ k:k\in\mathcal{U}_{n},\sum_{n^{'}=1}^{N}1\left(\left(k,n^{'}\right)\in\mathcal{E}\right)>1\right\} $
denote the set of all pico cell I-users in pico cell $n$. Then we
have $\mathcal{B}_{k}^{\textrm{MA}}=\mathcal{B}_{k^{'}}^{\textrm{MA}},\:\forall k,k^{'}\in\mathcal{U}_{n}^{I}$
and define $\mathcal{B}_{n}\triangleq\mathcal{B}_{k}^{\textrm{MA}},\forall k\in\mathcal{U}_{n}^{I}$
as the set of neighbor macro BSs of pico cell $n$.

\end{assumption}

The above assumption states that a macro BS will interfere with all
the I-users in a pico cell as long as it interferes with any user
in the pico cell. This is reasonable since the coverage area of a
macro BS is much larger than that of a pico BS.

\subsection{Structural Properties and Problem Transformation of $\mathcal{P}_{A}$\label{sub:Structural-Properties-ofPA}}

We exploit the interference structure in the HetNet topology $\mathcal{G}_{T}$
to derive the structural properties of $\mathcal{P}_{A}$. Throughout
this section, we will use the following example problem to illustrate
the intuition behind the main results.
\begin{example}
\label{exm-problemA}Consider $\mathcal{P}_{A}\left(\mathcal{G}_{T}\right)$
for the HetNet in Fig. \ref{fig:ABS_intro} with $N_{0}=2$ Macro
BSs and $N-N_{0}=1$ pico BS. The set of Type A users is $\mathcal{U}_{A}=\left\{ 1,2,3,4\right\} $
and the objective function is specified as $U_{A}\left(Q_{A},\pi_{A}\right)=\sum_{k=1}^{4}\overline{\mathcal{I}}_{m_{A},k}\left(Q_{A},\pi_{A}\right)$
(i.e., we consider sum-rate utility). For illustration, we focus on
the case when the marginal probability that a macro BS is transmitting
ABRB%
\footnote{The marginal probability that macro BS $n$ is transmitting ABRB is
$q_{n}^{A}=\sum_{\mathbf{a}\in\left\{ \mathbf{a}^{'}:\: a_{n}^{'}=0\right\} }Q_{A}\left(\mathbf{a}\right)$.%
} is fixed as $\mathbf{q}_{A}=\left[q_{1}^{A},q_{2}^{A}\right]^{T}=[0.7,0.5]^{T}$.\hfill \IEEEQED
\end{example}

Define two sets of ABRB patterns $\mathcal{A}_{n}$ and $\overline{\mathcal{A}}_{n}=\mathcal{A}/\mathcal{A}_{n}$
for each BS $n$. For macro BS, $\mathcal{A}_{n}$ is the set of ABRB
patterns under which macro BS $n$ is transmitting data. For pico
BS, $\mathcal{A}_{n}$ is the set of ABRB patterns under which all
of its neighbor macro BSs is transmitting ABRB. Using the configuration
in Example \ref{exm-problemA}, we have $\mathcal{A}_{1}=\left\{ \left[1,0\right],\left[1,1\right]\right\} $,
$\mathcal{A}_{2}=\left\{ \left[0,1\right],\left[1,1\right]\right\} $
and $\mathcal{A}_{3}=\left\{ \left[0,0\right],\left[1,0\right]\right\} $
(the formal definition of $\mathcal{A}_{n}$ and $\overline{\mathcal{A}}_{n}$
for general cases is in Appendix \ref{sub:SPAgeneral}).

In \textquotedbl{}Observation \ref{clm:Effect-of-ABRB-MI}\textquotedbl{},
we find that the ABRB pattern $\mathbf{a}$ only affects the data
rate of a Type A user in cell $n$ by whether $\mathbf{a}\in\mathcal{A}_{n}$
or not. Based on that, we find that the policy space for both the
ABRB control $Q_{A}$ and user scheduling $\pi_{A}$ can be significantly
reduced in \textquotedbl{}Observation \ref{clm:QAreduction} and \ref{clm:Policy-Space-ReductionpiA}\textquotedbl{}.
While these observations are made for the specific configuration in
Example \ref{exm-problemA}, they are also correct for general configurations
and are formally stated in Lemma \ref{lem:sysmAn}, Theorem \ref{thm:ABSreduce}
and Theorem \ref{thm:Policy-Space-ReductionpiA} in Appendix \ref{sub:SPAgeneral}.
Finally, using the above results, we transform the complicated problem
$\mathcal{P}_{A}$ to a much simpler problem with $\mathbf{q}_{A},\pi_{A}$
as the optimization variables.

\begin{observation}[Effect of ABRB on Mutual Information]\label{clm:Effect-of-ABRB-MI}For
given CSI $\mathbf{H}_{m_{A}}$ and a feasible user scheduling vector
$\rho_{m_{A}}^{n}$, the mutual information $\mathcal{I}_{m_{A},k}\left(\mathbf{a},\mathbf{H}_{m_{A}},\rho_{m_{A}}^{n}\right)$
of a Type A user $k$ in cell $n$ only depends on whether the ABRB
pattern $\mathbf{a}\in\mathcal{A}_{n}$ or not, i.e., $\mathcal{I}_{m_{A},k}\left(\mathbf{a},\mathbf{H}_{m_{A}},\rho_{m_{A}}^{n}\right)=\mathcal{I}_{m_{A},k}\left(\mathbf{a}^{'},\mathbf{H}_{m_{A}},\rho_{m_{A}}^{n}\right)$,$\forall\mathbf{a},\mathbf{a}^{'}\in\mathcal{A}_{n}$
(or $\forall\mathbf{a},\mathbf{a}^{'}\in\overline{\mathcal{A}}_{n}$).
Moreover, we have $\mathcal{I}_{m_{A},k}\left(\mathbf{a},\mathbf{H}_{m_{A}},\rho_{m_{A}}^{n}\right)\geq\mathcal{I}_{m_{A},k}\left(\mathbf{a}^{'},\mathbf{H}_{m_{A}},\rho_{m_{A}}^{n}\right)$
for any $\mathbf{a}\in\mathcal{A}_{n},\mathbf{a}^{'}\in\overline{\mathcal{A}}_{n}$.\end{observation}

Let us illustrate the above observation using the configuration in
Example \ref{exm-problemA}. There are 4 type A users: $\mathcal{U}_{A}=\left\{ 1,2,3,4\right\} $.
For user 1 in macro cell 1 (N-user), it is scheduled for transmission
whenever macro BS 1 is not transmitting ABRB (i.e., $a_{1}=1$). Hence
the mutual information is $\mathcal{I}_{m_{A},1}\left(\mathbf{a},\mathbf{H}_{m_{A}},\rho_{m_{A}}^{1}\right)=a_{1}\log\left(1+P_{1}\left|h_{m_{A},1,1}\right|^{2}\right)$,
which only depends on whether $a_{1}=1$ (i.e., $\mathbf{a}\in\mathcal{A}_{1}$)
or not. Moreover, the mutual information is higher if $\mathbf{a}\in\mathcal{A}_{1}$
because $\mathcal{I}_{m_{A},1}\left(\mathbf{a},\mathbf{H}_{m_{A}},\rho_{m_{A}}^{1}\right)=0,\forall\mathbf{a}\in\overline{\mathcal{A}}_{1}$.
For user 4 in pico cell 3 (I-user), if $\mathbf{a}\in\mathcal{A}_{3}$,
its neighbor macro BS 2 is transmitting ABRB and the mutual information
is $\mathcal{I}_{m_{A},4}\left(\mathbf{a},\mathbf{H}_{m_{A}},\rho_{m_{A}}^{3}\right)=\rho_{m,4}\log\left(1+P_{3}\left|h_{m_{A},4,3}\right|^{2}\right)$;
otherwise ($\mathbf{a}\in\overline{\mathcal{A}}_{3}$), macro BS 2
is transmitting data and we have $\rho_{m,4}=0$ and $\mathcal{I}_{m_{A},4}\left(\mathbf{a},\mathbf{H}_{m_{A}},\rho_{m_{A}}^{3}\right)=0$.
Similar observations can be made for user 2 and 3.

Based on Observation \ref{clm:Effect-of-ABRB-MI}, the ABRB control
policy space can be significantly reduced. 

\begin{figure}
\begin{centering}
\includegraphics[width=88mm]{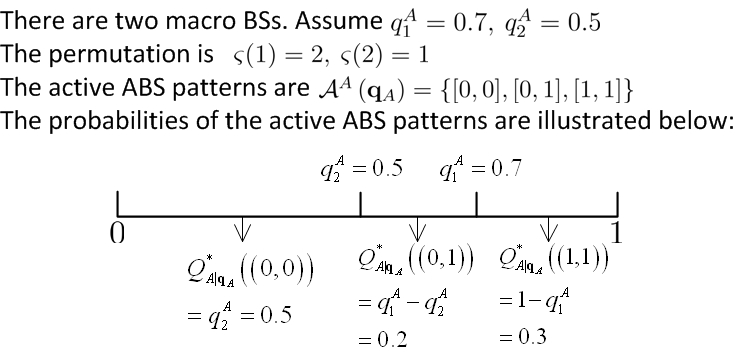}
\par\end{centering}

\caption{\label{fig:ABS_reduction}{\small Illustration of the structure of
the optimal $Q^{A}$ conditioned on a given $\mathbf{q}_{A}$}}
\end{figure}

\begin{observation}[Policy Space Reduction for $Q_{A}$]\label{clm:QAreduction}Consider
$\mathcal{P}_{A}\left(\mathcal{G}_{T}\right)$ for the configuration
in Example \ref{exm-problemA}. The optimal ABRB control of $\mathcal{P}_{A}\left(\mathcal{G}_{T}\right)$
conditioned on a given marginal probability vector $\mathbf{q}_{A}=[0.7,0.5]^{T}$,
denoted by $Q_{A|\mathbf{q}_{A}}^{*}$, has the \textit{synchronous
ABRB structure}%
\footnote{The formal definition of the synchronous ABRB structure for general
cases is in Theorem \ref{thm:ABSreduce}.%
} where the transmissions of ABRB at the macro BSs are aligned as much
as possible. As a result, there are only $3$ active ABRB patterns
$\left\{ [0,0],[0,1],[1,1]\right\} $ and the corresponding pmf $Q_{A|\mathbf{q}_{A}}^{*}\left(\cdot\right)$
is given by a function of $\mathbf{q}_{A}$ as illustrated in Fig.
\ref{fig:ABS_reduction}.\end{observation}

In general, for a HetNet with $N_{0}$ macro BSs, there are only $N_{0}+1$
active ABRB patterns under synchronous ABRB structure, which is significantly
smaller than the number of all possible ABRB patterns $2^{N_{0}}$.
As a result, the optimization $\mathcal{P}_{A}\left(\mathcal{G}_{T}\right)$
w.r.t $Q_{A}$ (with $2^{N_{0}}$ variables) can be reduced to an
equivalent optimization w.r.t. $\mathbf{q}_{A}$ (with $N_{0}$ variables)
with much lower dimensions. Observation \ref{clm:QAreduction} can
be understood as follows. By Observation \ref{clm:Effect-of-ABRB-MI},
a higher average mutual information can be achieved for user $k\in\mathcal{U}_{n}$
under the ABRB patterns $\mathbf{a}\in\mathcal{A}_{n}$. Hence, for
given marginal probabilities $\mathbf{q}_{A}=[0.7,0.5]^{T}$, the
average mutual information region will be maximized if we can simultaneously
maximize $\sum_{\mathbf{a}\in\mathcal{A}_{n}}Q_{A}\left(\mathbf{a}\right)$
for all BSs $1\leq n\leq3$. For macro BSs $n=1,2$, we have $\sum_{\mathbf{a}\in\mathcal{A}_{n}}Q_{A}\left(\mathbf{a}\right)=q_{n}^{A}$,
which is fixed for given $\mathbf{q}_{A}$. For pico BSs $n=3$, we
have $\sum_{\mathbf{a}\in\mathcal{A}_{n}}Q_{A}\left(\mathbf{a}\right)\leq\underset{j\in\mathcal{B}_{n}=\left\{ 1,2\right\} }{\textrm{min}}\left(q_{j}^{A}\right)=0.5$,
and the equality holds if and only if $Q_{A}$ has the synchronous
ABRB structure in Fig. \ref{fig:ABS_reduction}.

Similarly, we can reduce the user scheduling policy space using Observation
\ref{clm:Effect-of-ABRB-MI}.

\begin{observation}[Policy Space Reduction for $\pi_{A}$]\label{clm:Policy-Space-ReductionpiA}Consider
$\mathcal{P}_{A}\left(\mathcal{G}_{T}\right)$ for the configuration
in Example \ref{exm-problemA}. For given CSI $\mathbf{H}_{m_{A}}$
and ABRB pattern $\mathbf{a}$, the optimal user scheduling at BS
$n$ is given by{\small 
\begin{equation}
\pi_{A}^{n}\left(\mathbf{a},\mathbf{H}_{m_{A}}\right)=\underset{\rho_{m_{A}}^{n}\in\Gamma_{m_{A}}^{n}\left(\mathbf{a}\right)}{\textrm{argmax}}\sum_{k\in\mathcal{U}_{A}\cap\mathcal{U}_{n}}\mathcal{I}_{m_{A},k}\left(\mathbf{a},\mathbf{H}_{m_{A}},\rho_{m_{A}}^{n}\right).\label{eq:argmaxruo}
\end{equation}
}By Observation \ref{clm:Effect-of-ABRB-MI}, if $\rho_{m_{A}}^{n*}$
solves the maximization problem (\ref{eq:argmaxruo}) for certain
$\mathbf{a}\in\mathcal{A}_{n}$, it solves (\ref{eq:argmaxruo}) for
all $\mathbf{a}\in\mathcal{A}_{n}$. Hence, it will not loss optimality
by imposing an additional constraint on the user scheduling such that
$\pi_{A}^{n}\left(\mathbf{a},\mathbf{H}_{m_{A}}\right)=\pi_{A}^{n}\left(\mathbf{a}^{'},\mathbf{H}_{m_{A}}\right),\forall\mathbf{a},\mathbf{a}^{'}\in\mathcal{A}_{n}$
(or $\forall\mathbf{a},\mathbf{a}^{'}\in\overline{\mathcal{A}}_{n}$).\end{observation}

For convenience, let $\Xi_{A}^{*}$ denote the set of all feasible
user scheduling policies satisfying the above constraint in Observation
\ref{clm:Policy-Space-ReductionpiA} (The formal definition of $\Xi_{A}^{*}$
is given in Theorem \ref{thm:Policy-Space-ReductionpiA}). Then for
given $\mathbf{q}_{A}$, $\pi_{A}\in\Xi_{A}^{*}$ and under the synchronous
ABRB, the corresponding objective function of $\mathcal{P}_{A}$ can
be rewritten as $U_{A}\left(Q_{A|\mathbf{q}_{A}}^{*},\pi_{A}\right)=\hat{U}_{A}\left(\mathbf{q}_{A},\pi_{A}\right)$,
where
\begin{equation}
\hat{U}_{A}\left(\mathbf{q}_{A},\pi_{A}\right)=\sum_{k\in\mathcal{U}_{A}}w_{k}u\left(\overline{\mathcal{I}}_{m_{A},k}\left(Q_{A|\mathbf{q}_{A}}^{*},\pi_{A}\right)\right),\label{eq:UAobj}
\end{equation}
and $\overline{\mathcal{I}}_{m_{A},k}\left(Q_{A|\mathbf{q}_{A}}^{*},\pi_{A}\right)$
is the corresponding average mutual information given in (\ref{eq:IBARAk})
of Appendix \ref{sub:SPAgeneral}. As a result, the subproblem $\mathcal{P}_{A}\left(\mathcal{G}_{T}\right)$
can be transformed into a simpler problem with $\mathcal{O}\left(N_{0}\right)$
solution space.
\begin{cor}
[Equivalent Problem Transformation of $\mathcal{P}_{A}\left(\mathcal{G}_{T}\right)$]\label{cor:Equivalent-Problem-PA}Let
$\mathbf{q}_{A}^{*},\pi_{A}^{*}$ denote the optimal solution of the
following joint optimization problem.
\begin{equation}
\max_{\mathbf{q}_{A},\pi_{A}}\:\hat{U}_{A}\left(\mathbf{q}_{A},\pi_{A}\right),\:\textrm{s.t.}\:\mathbf{q}_{A}\in\mathcal{Q}^{A},\pi_{A}\in\Xi_{A}^{*},\label{eq:PLAre}
\end{equation}
where $\mathcal{Q}^{A}=\left\{ \mathbf{q}_{A}:\: q_{j}^{A}\in\left[0,1\right],\forall j\right\} $.
Then $Q_{A|\mathbf{q}_{A}^{*}}^{*}$, $\pi_{A}^{*}$ is the optimal
solution of problem $\mathcal{P}_{A}\left(\mathcal{G}_{T}\right)$.

Furthermore, problem (\ref{eq:PLAre}) is a bi-convex problem, i.e.,
for fixed $\pi_{A}$, problem (\ref{eq:PLAre}) is convex w.r.t. $\mathbf{q}_{A}$,
and for fixed $\mathbf{q}_{A}$, problem (\ref{eq:PLAre}) is also
convex w.r.t. $\pi_{A}$.
\end{cor}

Please refer to Appendix \ref{sub:Proof-of-CorollayPA} for the proof.

\subsection{Two Timescale Alternating Optimization Algorithm for $\mathcal{P}_{A}$\label{sub:Two-Timescale-AlternatingPA}}

By Corollary \ref{cor:Equivalent-Problem-PA}, we only need to solve
the equivalent problem of $\mathcal{P}_{A}$ in (\ref{eq:PLAre}).
Since problem (\ref{eq:PLAre}) is bi-convex, we propose the following
\textit{Two Timescale Alternating Optimization (AO)} algorithm. For
notation convenience, time index $t$ and $T$ are used to denote
the subframe index and super-frame index respectively, where a super-frame
consists of $L_{S}$ subframes. \smallskip{}

\textit{Algorithm AO\_A} (Two Timescale AO for $\mathcal{P}_{A}\left(\mathcal{G}_{T}\right)$): 

\textbf{\small Initialization}{\small : Choose proper initial $\mathbf{q}_{A}^{(0)}$,$\pi_{A}^{(-1)}$.
Set $T=0$.}{\small \par}

\textbf{\small Step 1}{\small{} (Short timescale user scheduling optimization):
For fixed $\mathbf{q}_{A}^{\left(T\right)}$, let $\pi_{A}^{(T)}=\left\{ \pi_{A}^{n(T)},n=1,...,N\right\} $,
where $\pi_{A}^{n(T)}$ is given by
\begin{eqnarray*}
\pi_{A}^{n(T)}\left(\mathbf{a},\mathbf{H}_{m_{A}}\right)=\;\;\;\;\;\;\;\;\;\;\;\;\;\;\;\;\;\;\;\;\;\;\;\;\;\;\;\;\;\;\;\;\;\;\;\;\;\;\;\;\;\;\;\;\;\;\;\;\;\;\;\;\;\;\;\;\;\;\;\;\;\;\;\;\;\;\;\;\\
\underset{\rho_{m_{A}}^{n}\in\Gamma_{m_{A}}^{n}\left(\mathbf{a}\right)}{\textrm{argmax}}\sum_{k\in\mathcal{U}_{A}\cap\mathcal{U}_{n}}w_{k}\frac{\partial u\left(r\right)}{\partial r}|_{r=\overline{r}_{k}^{(T)}}\mathcal{I}_{m_{A},k}\left(\mathbf{a},\mathbf{H}_{m_{A}},\rho_{m_{A}}^{n}\right),
\end{eqnarray*}
where $\overline{r}_{k}^{(T)}=q_{s}M\overline{\mathcal{I}}_{m_{A},k}\left(Q_{A|\mathbf{q}_{A}^{\left(T\right)}}^{*},\pi_{A}^{(T)}\right)$
is the average data rate of user $k$ under $\mathbf{q}_{A}^{\left(T\right)}$
and user scheduling policy $\pi_{A}^{(T)}$. For each subframe $t\in\left[TL_{S},\left(T+1\right)L_{S}-1\right]$,
the user scheduling vector of BS $n$ is given by $\rho_{m_{A}}^{n}=\pi_{A}^{n(T)}\left(\mathbf{a}_{m_{A}},\mathbf{H}_{m_{A}}\right)$,
where $\mathbf{H}_{m_{A}}$ and $\mathbf{a}_{m_{A}}$ are the CSI
and ABRB pattern at the $t$-th subframe.}{\small \par}

\textbf{\small Step 2 }{\small (Long timescale ABRB optimization):}\textbf{\small{}
}{\small Find the optimal solution $\mathbf{q}_{A}^{(T+1)}$ of problem
(\ref{eq:PLAre}) under fixed $\pi_{A}^{(T)}$ using e.g., Ellipsoid
method. Let $T=T+1$.}{\small \par}

\textbf{\small Return to Step 1 until $\mathbf{q}_{A}^{(T)}=\mathbf{q}_{A}^{(T-1)}$}{\small{}
or the maximum number of iterations is reached.}\smallskip{}

While (\ref{eq:PLAre}) is a bi-convex problem and AO algorithm is
known to converge to local optimal solutions only, we exploit the
hidden convexity of the problem and show below that Algorithm AO\_A
can converge to the global optimal solution under certain conditions.
\begin{thm}
[Global Convergence of Algorithm AO\_A]\label{thm:Global-Convergence-LA}Let
$\left[\mathbf{q}_{A}^{(T+1)},\pi_{A}^{(T)}\right]=\mathcal{F}\left(\left[\mathbf{q}_{A}^{(T)},\pi_{A}^{(T-1)}\right]\right)$
denote the iterate sequence of Algorithm AO\_A began at $\mathbf{q}_{A}^{(0)},\pi_{A}^{(-1)}$,
and denote the set of fixed points of the mapping $\mathcal{F}$ as
$\Delta=\left\{ \mathbf{q}_{A}\in\mathcal{Q}^{A},\pi_{A}\in\Xi_{A}^{*}:\:\left[\mathbf{q}_{A},\pi_{A}\right]\right.$
$\left.=\mathcal{F}\left(\left[\mathbf{q}_{A},\pi_{A}\right]\right)\right\} $.
For any $\mathbf{q}_{A}\in\mathcal{Q}^{A},\pi_{A}\in\Xi_{A}^{*}$
that is not a fixed point of $\mathcal{F}$, assume that $\hat{U}_{A}\left(\mathbf{q}_{A}^{'},\pi_{A}^{'}\right)\neq\hat{U}_{A}\left(\mathbf{q}_{A},\pi_{A}\right)$,
where $\left[\mathbf{q}_{A}^{'},\pi_{A}^{'}\right]=\mathcal{F}\left(\left[\mathbf{q}_{A},\pi_{A}\right]\right)$.
Then:
\begin{enumerate}
\item Algorithm AO\_A converges to a fixed point $\left[\tilde{\mathbf{q}}_{A},\tilde{\pi}_{A}\right]\in\Delta$
of $\mathcal{F}$.
\item Any fixed point $\left[\tilde{\mathbf{q}}_{A},\tilde{\pi}_{A}\right]\in\Delta$
is a globally optimal solution of problem (\ref{eq:PLAre}).
\end{enumerate}
\end{thm}

Please refer to Appendix \ref{sub:Proof-of-TheoremLA} for the proof.

Step 1 of Algorithm AO\_A requires the knowledge of the average data
rate $\overline{r}_{k}^{(T)}$ under $\mathbf{q}_{A}^{\left(T\right)}$
and $\pi_{A}^{(T)}$. We adopt a reasonable approximation on $\overline{r}_{k}^{(T)}$
using a moving average data rate $R_{k}\left(t\right),\forall t\in\left[TL_{S},\left(T+1\right)L_{S}-1\right]$
given by \cite{Whiting_TWC04_PFSconv}
\begin{equation}
R_{k}\left(t\right)=\frac{t-TL_{S}+1}{t-TL_{S}+2}R_{k}\left(t-1\right)+\frac{1}{t-TL_{S}+2}r_{k}\left(t-1\right),\label{eq:RbarAppr}
\end{equation}
where $r_{k}\left(t\right)$ is the data rate delivered to user $k$
at subframe $t$.
\begin{rem}
If we replace $\overline{r}_{k}^{(T)}$ in step 1 of Algorithm AO\_A
with the approximation $R_{k}\left(t\right)$ in (\ref{eq:RbarAppr}),
the global convergence result in Theorem \ref{thm:Global-Convergence-LA}
no longer holds. However, it has been shown in \cite{Whiting_TWC04_PFSconv}
that $R_{k}\left(t\right)$ converges to $\overline{r}_{k}^{(T)}$
as $t-TL_{S}\rightarrow\infty$. Hence, with the approximation $R_{k}\left(t\right)$,
Algorithm AO\_A is still asymptotically optimal for large super frame
length $L_{S}$.
\end{rem}

In step 2 of Algorithm AO\_A, the average mutual information $\overline{\mathcal{I}}_{m_{A},k}\left(Q_{A|\mathbf{q}_{A}^{(T)}}^{*},\pi_{A}^{(T)}\right)$
in the optimization objective contains two intermediate problem parameters
$e_{k}^{A}\left(\pi_{A}^{(T)}\right)$ and $\overline{e}_{k}^{A}\left(\pi_{A}^{(T)}\right)$
defined under (\ref{eq:IBARAk}) in Appendix \ref{sub:SPAgeneral}.
The calculation of $e_{k}^{A}\left(\pi_{A}^{(T)}\right)$'s and $\overline{e}_{k}^{A}\left(\pi_{A}^{(T)}\right)$'s
requires the knowledge of the distribution of all the channel coefficients,
which is usually difficult to obtain offline. However, these terms
can be easily estimated online using the time average of the sampled
data rates delivered to user $k$ under ABRB patterns $\mathcal{A}_{b_{k}}$
and $\overline{\mathcal{A}}_{b_{k}}$ respectively. The ABRB control
$\mathbf{q}_{A}^{(T+1)}$ is then obtained by solving the long timescale
problem in step 2 with say, the ellipsoid method based on these estimates.

\section{Two Timescale Hierarchical Solution for Co-Tier Interference Control
(Subproblem B) \label{sub:Solution-of-SubproblemPB}}

The number of ABRB control variables $Q_{B}\left(\mathbf{a}\right),\forall\mathbf{a}\in\mathcal{A}$
is exponentially large w.r.t. $N_{0}$. To simplify $\mathcal{P}_{B}$,
we introduce an auxiliary variable called the \textit{ABRB profile}
and decompose $\mathcal{P}_{B}$.

\subsection{Problem Decomposition of $\mathcal{P}_{B}$}

We first define the ABRB profile.
\begin{defn}
[ABRB Profile] The ABRB profile $\mathcal{A}^{B}\subset\mathcal{A}$
is a subset of ABRB patterns for Type B subbands.\hfill \IEEEQED
\end{defn}

Using the notion of ABRB Profile and primal decomposition, $\mathcal{P}_{B}\left(\mathcal{G}_{T}\right)$
can be approximated by two subproblems:

\textbf{Optimization of $Q_{B}$ and $\pi_{B}$ for a given ABRB Profile
$\mathcal{A}^{B}$.}
\begin{eqnarray*}
 & \mathcal{P}_{B}^{I}\left(\mathcal{G}_{T};\mathcal{A}^{B}\right): & U_{B}^{I}\left(\mathcal{A}^{B}\right)\triangleq\max_{Q_{B},\pi_{B}}U_{B}\left(Q_{B},\pi_{B}\right),\\
 & \textrm{s.t.} & \sum_{\mathbf{a}\in\mathcal{A}^{B}}Q_{B}\left(\mathbf{a}\right)=1;\: Q_{B}\left(\mathbf{a}\right)\geq0,\forall\mathbf{a}\in\mathcal{A}^{B},\\
 &  & Q_{B}\left(\mathbf{a}\right)=0,\:\forall\mathbf{a}\notin\mathcal{A}^{B}.
\end{eqnarray*}

\textbf{Optimization of ABRB profile $\mathcal{A}^{B}$.}
\[
\mathcal{P}_{B}^{II}\left(\mathcal{G}_{T}\right):\:\max_{\mathcal{A}^{B}}U_{B}^{I}\left(\mathcal{A}^{B}\right),\:\textrm{s.t.}\:\mathcal{A}^{B}\subset\mathcal{A};\:\left|\mathcal{A}^{B}\right|\leq\left|\mathcal{U}_{B}\right|.
\]

In $\mathcal{P}_{B}^{II}\left(\mathcal{G}_{T}\right)$, we restrict
the size of $\mathcal{A}^{B}$ to be no more than $\left|\mathcal{U}_{B}\right|$.
In Appendix \ref{sub:Proof-of-TheoremEquPB}, we prove that at the
asymptotically optimal solution of $\mathcal{P}_{B}\left(\mathcal{G}_{T}\right)$
as SNR becomes high, the number of active ABRB patterns is indeed
less than or equal to $\left|\mathcal{U}_{B}\right|$.

\subsection{Two Timescale Alternating Optimization Algorithm for $\mathcal{P}_{B}^{I}$}

Let $\mathbf{a}_{j}^{B}$ denote the $j^{\textrm{th}}$ ABRB pattern
in $\mathcal{A}^{B}$. Then the average mutual information $\overline{\mathcal{I}}_{m_{B},k}$
can be rewritten as
\begin{equation}
\overline{\mathcal{I}}_{m_{B},k}\left(\mathbf{q}_{B},\pi_{B}\right)=\sum_{j=1}^{\left|\mathcal{A}^{B}\right|}q_{j}^{B}I_{m_{B},k}\left(\pi_{B},\mathbf{a}_{j}^{B}\right),\label{eq:IbarB}
\end{equation}
where $\mathbf{q}^{B}=\left[q_{1}^{B},...,q_{\left|\mathcal{A}^{B}\right|}^{B}\right]$
with $q_{j}^{B}=Q_{B}\left(\mathbf{a}_{j}^{B}\right)$ denoting the
probability that the ABRB pattern $\mathbf{a}_{j}^{B}$ is used. Hence
$\mathcal{P}_{B}^{I}\left(\mathcal{G}_{T};\mathcal{A}^{B}\right)$
can be reformulated as 

\begin{eqnarray}
\max_{\mathbf{q}_{B},\pi_{B}} & \hat{U}_{B}\left(\mathbf{q}_{B},\pi_{B}\right)\triangleq\sum_{k\in\mathcal{U}_{B}}w_{k}u\left(\overline{\mathcal{I}}_{m_{B},k}\left(\mathbf{q}_{B},\pi_{B}\right)\right),\label{eq:PLBequ}\\
 & \textrm{s.t.}\:\sum_{j=1}^{\left|\mathcal{A}^{B}\right|}q_{j}^{B}=1;\:\textrm{and}\: q_{j}^{B}\geq0,\forall j.\nonumber 
\end{eqnarray}

Similar to (\ref{eq:PLAre}), we propose a two timescale AO to solve
for problem (\ref{eq:PLBequ}) w.r.t. $\pi_{B}$ and $\mathbf{q}_{B}$.\smallskip{}

\textit{Algorithm AO\_B} (Two timescale AO for $\mathcal{P}_{B}^{I}\left(\mathcal{G}_{T};\mathcal{A}^{B}\right)$): 

\textbf{\small Initialization}{\small : Choose proper initial $\mathbf{q}_{B}^{(0)}$,$\pi_{B}^{(-1)}$.
Set $T=0$.}{\small \par}

\textbf{\small Step 1}{\small{} (Short timescale user scheduling optimization):
For fixed $\mathbf{q}_{B}^{\left(T\right)}$, let $\pi_{B}^{(T)}=\left\{ \pi_{B}^{n(T)},n=1,...,N\right\} $,
where $\pi_{B}^{n(T)}$ is given by
\begin{eqnarray*}
\pi_{B}^{n(T)}\left(\mathbf{a},\mathbf{H}_{m_{B}}\right)=\;\;\;\;\;\;\;\;\;\;\;\;\;\;\;\;\;\;\;\;\;\;\;\;\;\;\;\;\;\;\;\;\;\;\;\;\;\;\;\;\;\;\;\;\;\;\;\;\;\;\;\;\;\;\;\;\;\;\;\;\;\;\;\;\;\;\;\;\\
\underset{\rho_{m_{B}}^{n}\in\Gamma_{m_{B}}^{n}\left(\mathbf{a}\right)}{\textrm{argmax}}\sum_{k\in\mathcal{U}_{B}\cap\mathcal{U}_{n}}w_{k}\frac{\partial u\left(r\right)}{\partial r}|_{r=\overline{r}_{k}^{(T)}}\mathcal{I}_{m_{B},k}\left(\mathbf{a},\mathbf{H}_{m_{B}},\rho_{m_{B}}^{n}\right),
\end{eqnarray*}
where $\overline{r}_{k}^{(T)}=\left(1-q_{s}\right)M\overline{\mathcal{I}}_{m_{B},k}\left(\mathbf{q}_{B}^{(T)},\pi_{B}^{(T)}\right)$
is the average data rate of user $k$ under $\mathbf{q}_{B}^{\left(T\right)}$
and user scheduling policy $\pi_{B}^{(T)}$. For each subframe $t\in\left[TL_{S},\left(T+1\right)L_{S}-1\right]$,
the user scheduling vector of BS $n$ is given by $\rho_{m_{B}}^{n}=\pi_{B}^{n(T)}\left(\mathbf{a}_{m_{B}},\mathbf{H}_{m_{B}}\right)$,
where $\mathbf{H}_{m_{B}}$ and $\mathbf{a}_{m_{B}}$ are the CSI
and ABRB pattern at the $t$-th subframe.}{\small \par}

\textbf{\small Step 2 }{\small (Long timescale ABRB optimization):}\textbf{\small{}
}{\small Find the optimal solution $\mathbf{q}_{B}^{(T+1)}$ of problem
(\ref{eq:PLBequ}) under fixed $I_{m_{B},k}\left(\pi_{B}^{(T)},\mathbf{a}_{j}^{B}\right)$'s
using e.g., interior point method. Let $T=T+1$.}{\small \par}

\textbf{\small Return to Step 1 until $\mathbf{q}_{B}^{(T)}=\mathbf{q}_{B}^{(T-1)}$}{\small{}
or the maximum number of iterations is reached.}{\small \par}

\smallskip{}

Using similar proof as that in Appendix \ref{sub:Proof-of-TheoremLA},
it can be shown that Algorithm AO\_B converges to the global optimal
solution for problem (\ref{eq:PLBequ}) under similar assumption as
in Theorem \ref{thm:Global-Convergence-LA}. The average data rate
$\overline{r}_{k}^{(T)}$ and the mutual information $I_{m_{B},k}\left(\pi_{B}^{(T)},\mathbf{a}_{j}^{B}\right)$
can be estimated online in a similar way as in Algorithm AO\_A.

\subsection{Finding a Good ABRB Profile for $\mathcal{P}_{B}^{II}\left(\mathcal{G}_{T}\right)$}

Problem $\mathcal{P}_{B}^{II}\left(\mathcal{G}_{T}\right)$ is a difficult
combinatorial problem and the complexity of finding the optimal ABRB
Profile is extremely high. In this section, we illustrate the top
level method for finding a good ABRB profile. The detailed algorithm
to solve $\mathcal{P}_{B}^{II}\left(\mathcal{G}_{T}\right)$ is given
in Appendix \ref{sub:Proof-of-TheoremEquPB}.

A good ABRB profile can be found based on an \textit{interference
graph} extracted from the topology graph.
\begin{defn}
[Interference Graph]\label{def:IGraph}For a HetNet Topology Graph
$\mathcal{G}_{T}=\left\{ \mathcal{B},\mathcal{U},\mathcal{E}\right\} $,
define an undirected interference graph $\mathcal{G}_{I}\left(\mathcal{G}_{T}\right)=\left\{ \mathcal{L},\mathcal{E}_{I}\right\} $,
where $\mathcal{L}=\left\{ l_{k},\forall k\in\mathcal{U}_{B}\right\} $
is the vertex set and $\mathcal{E}_{I}=\left\{ e\left(l_{k},l_{k^{'}}\right)\in\left\{ 0,1\right\} ,\forall k\neq k^{'}\in\mathcal{U}_{B}\right\} $
is edge set with $e\left(l_{k},l_{k^{'}}\right)$ denoting the edge
between $l_{k}$ and $l_{k^{'}}$. For any $k\neq k^{'}\in\mathcal{U}_{B}$,
if $\left\{ \left(k,b_{k^{'}}\right)\cup\left(k^{'},b_{k}\right)\right\} \cap\mathcal{E}\neq\Phi$,
$e\left(l_{k},l_{k^{'}}\right)=1$, otherwise, $e\left(l_{k},l_{k^{'}}\right)=0$,
where $\Phi$ is the void set.\hfill \IEEEQED
\end{defn}

\begin{figure}
\begin{centering}
\includegraphics[width=88mm]{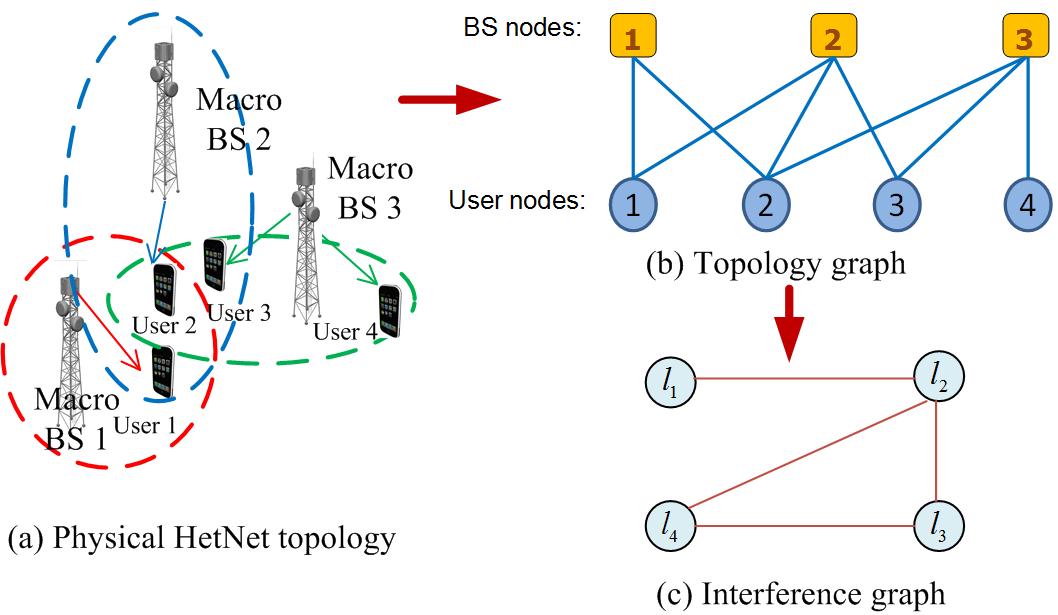}
\par\end{centering}

\caption{\label{fig:IFC_graph1}{\small An example of extracting the interference
graph from the HetNet topology graph}}
\end{figure}

Fig. \ref{fig:IFC_graph1} illustrates how to extract the interference
graph from the topology graph using an example HetNet. Given an interference
graph $\mathcal{G}_{I}\left(\mathcal{G}_{T}\right)$ for the HetNet,
any two links $l_{k},l_{k^{'}}$ having an edge (i.e., $e\left(l_{k},l_{k^{'}}\right)=1$)
should not be scheduled for transmission simultaneously. On the other
hand, we should ``turn on'' as many ``non-conflicting'' links
as possible to maximize the spatial reuse efficiency. This intuition
suggests that the optimal ABRB profile is highly related to the \textit{maximal
independent set} of the interference graph $\mathcal{G}_{I}\left(\mathcal{G}_{T}\right)$.

\begin{defn}
[Maximal Independent Set (MIS)]A subset $\mathcal{V}$ of $\mathcal{L}$
is an \textit{independent set} of $\mathcal{G}_{I}\left(\mathcal{G}_{T}\right)=\left\{ \mathcal{L},\mathcal{E}_{I}\right\} $
if $e\left(l_{k},l_{k^{'}}\right)=0,\:\forall l_{k},l_{k^{'}}\in\mathcal{V},\:\textrm{and}\: l_{k}\neq l_{k^{'}}$.
A \textit{maximal independent set} (MIS) is an independent set that
is not a proper subset of any other independent set. For any MIS $\mathcal{V}$,
define $\mathcal{N}\left(\mathcal{V}\right)=\left\{ n:\:\exists k\in\mathcal{U}_{B}\cap\mathcal{U}_{n},\:\textrm{s.t.}\: l_{k}\in\mathcal{V}\right\} $
as the \textit{maximal independent macro BS set} corresponding to
$\mathcal{V}$. Let $\Theta_{T}\left(\mathcal{G}_{T}\right)$ denote
the set of all MISs of $\mathcal{G}_{I}\left(\mathcal{G}_{T}\right)$.\hfill \IEEEQED 
\end{defn}

For example, the set of all MISs of the interference graph in Fig.
\ref{fig:IFC_graph1} is $\Theta_{T}=\left\{ \mathcal{V}_{1}\triangleq\left\{ l_{1},l_{3}\right\} ,\mathcal{V}_{2}\triangleq\left\{ l_{1},l_{4}\right\} ,\mathcal{V}_{3}\triangleq\left\{ l_{2}\right\} \right\} $.
Define a set 
\[
\Psi\triangleq\left\{ \Theta=\left\{ \mathcal{V}_{1},...,\mathcal{V}_{\left|\Theta\right|}\right\} :\:\cup_{j=1}^{\left|\Theta\right|}\mathcal{V}_{j}=\mathcal{L}\right\} .
\]
Then the top level flow of finding a good ABRB profile is summarized
in Table. \ref{tab:flowPB}. In Step 2, we need to find a set of MISs
$\Theta^{*}\in\Psi$. For the example in Fig. \ref{fig:IFC_graph1},
$\Psi$ has a unique element given by $\Theta_{T}$ and thus $\Theta^{*}=\Theta_{T}$.
In Step 3, the mapping from $\Theta^{*}$ to the ABRB profile $\mathcal{A}^{B*}$
is $\mathcal{A}^{B*}=\left\{ \mathbf{a}^{B*}(j),\: j=1,...,\left|\Theta^{*}\right|\right\} $,
where $\mathbf{a}^{B*}(j)=\left[a_{1}^{B*}(j),...,a_{N_{0}}^{B*}(j)\right]$
with $a_{n}^{B*}(j)=1,\:\forall n\in\mathcal{N}\left(\mathcal{V}_{j}^{*}\right)$
and $a_{n}^{B*}(j)=0,\:\forall n\notin\mathcal{N}\left(\mathcal{V}_{j}^{*}\right)$.
For example, for the HetNet in Fig. \ref{fig:IFC_graph1}, we have
$\Theta^{*}=\Theta_{T}$ and thus $\mathcal{A}^{B*}=\left\{ \mathbf{a}^{B*}(1),\mathbf{a}^{B*}(2),\mathbf{a}^{B*}(3)\right\} $,
where $\mathbf{a}^{B*}(1)=\left[1,0,1\right]$, $\mathbf{a}^{B*}(2)=\left[1,0,1\right]$
and $\mathbf{a}^{B*}(3)=\left[0,1,0\right]$. Since $\mathbf{a}^{B*}(1)=\mathbf{a}^{B*}(2)$
in this case, $\mathcal{A}^{B*}$ can be reduced to $\mathcal{A}^{B*}=\left\{ \left[1,0,1\right],\left[0,1,0\right]\right\} $.

\begin{table}
\caption{\label{tab:flowPB}Top level flow of finding a good ABRB profile}

\centering{}%
\begin{tabular}{l}
\hline 
\textbf{\small Step 1: }Extract the interference graph from the \tabularnewline
{\small $\;$$\;$$\;$$\;$}topology graph using Definition \ref{def:IGraph}.\tabularnewline
\textbf{\small Step 2: }Find a set of MISs $\Theta^{*}\in\Psi$ using
Algorithm B2.\tabularnewline
\textbf{\small Step 3: }Obtain the ABRB profile $\mathcal{A}^{B*}$
by a mapping from $\Theta^{*}$.\tabularnewline
\hline 
\end{tabular}
\end{table}

For a general HetNet, $\Psi$ may have multiple elements and we need
to find a $\Theta^{*}\in\Psi$ such that the corresponding $\mathcal{A}^{B*}$
is a good solution of $\mathcal{P}_{B}^{II}\left(\mathcal{G}_{T}\right)$.
The detailed algorithm (Algorithm B2) for finding such $\Theta^{*}$
is given in Appendix \ref{sub:Proof-of-TheoremEquPB}, where we also
show that the $\mathcal{A}^{B*}$ found by the proposed algorithm
is asymptotically optimal for $\mathcal{P}_{B}^{II}\left(\mathcal{G}_{T}\right)$.

\section{Results and Discussions\label{sub:overall_online_Alg}}

\subsection{Summary of the Overall Solution and Implementation}

\begin{figure}
\begin{centering}
\includegraphics[width=88mm]{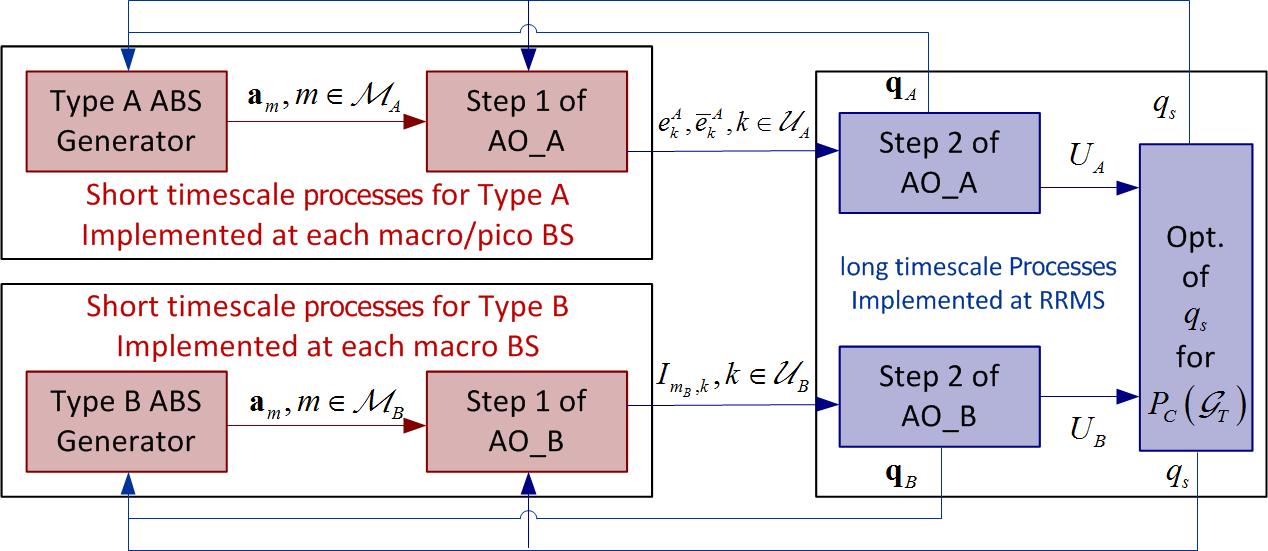}
\par\end{centering}

\caption{\label{fig:moduleconn}{\small Summary of overall solution and the
inter-relationship of the algorithm components. The red / blue blocks
represent long timescale / short timescale processes. The red / blue
arrows represent long-term / short-term signaling.}}
\end{figure}

Fig. \ref{fig:moduleconn} summarizes the overall solution and the
inter-relationship of the algorithm components for the Hierarchical
RRM. The solutions are divided into \textit{long timescale process}
and \textit{short timescale process}. The long timescale processing
consists of the step 2 of Algorithm AO\_A and AO\_B as well as the
sub-band partitioning $\mathcal{P}_{C}\left(\mathcal{G}_{T}\right)$.
The short timescale processing consists of step 1 of Algorithm AO\_A
and AO\_B as well as the generation of ABRB patterns. All the long
timescale processes are implemented globally at the RRMS and all the
short timescale processes are implemented locally at each macro and
pico BS as illustrated in Fig. \ref{fig:moduleconn}. At each super-frame
($L_{S}$ subframes), $\left(q_{s},\mathbf{q}_{A},\mathbf{q}_{B}\right)$
are computed from the RRMS and pass to the macro and pico BSs. Locally
at each BS and each subframe, the ABRB patterns on Type A subbands
$\mathbf{a}_{m},m\in\mathcal{M}_{A}$ and that on Type B subbands
$\mathbf{a}_{m},m\in\mathcal{M}_{B}$ are generated from the distribution
$\mathbf{q}_{A}$ and $\mathbf{q}_{B}$ respectively. Furthermore,
at each subframe, the user scheduling $\pi_{A}^{n}$ and $\pi_{B}^{n}$
are determined at each BS $n$ based on the instantaneous channel
quality indicator (CQI) of the direct links from the BS to the users.
At the end of the $L_{S}$ subframes, the macro and pico BSs deliver
the estimates of the data rates $\left\{ e_{k}^{A}\left(\pi_{A}\right),\overline{e}_{k}^{A}\left(\pi_{A}\right),I_{m_{B},k}\left(\pi_{B},\mathbf{a}_{j}^{B}\right)\right\} $
to the RRMS.

There are several advantages of the proposed hierarchical RRM. For
example, each BS only requires the direct link CQI. Hence, this solution
has low signaling overhead and good scalability on the complexity.
Furthermore, only long term statistical information is needed at the
RRMS. Hence, the solution is robust w.r.t. backhaul latency.

\subsection{Simulation Performance}

\begin{figure}
\begin{centering}
\includegraphics[scale=0.28]{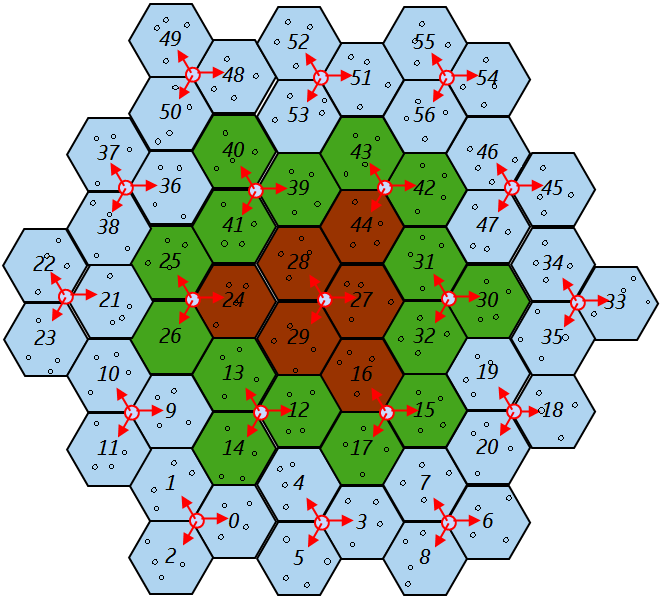}
\par\end{centering}

\caption{\label{fig:HetNet-topo}{\small Topology of a HetNet. Each red circle
stands for one eNB. The red arrow stands for a directional macro BS
and the corresponding hexagon is its coverage. The small black circles
stand for the pico BSs.}}
\end{figure}

In this section, we consider a HetNet with $19$ eNBs and $57$ directional
macro cells (three 120 degree sectors), as illustrated in Fig. \ref{fig:HetNet-topo}.
In each macro cell, there are $4$ uniformly distributed pico BSs.
There are $30$ users in one macro cell, $2/3$ of whom are clustered
around the pico BSs, while others are uniformly distributed within
the macro cell. The macro cells are separated in $500$ meters, and
the maximum transmit power of macro BSs and pico BSs are $46$ dBm
and $30$ dBm, respectively. The PFS utility is considered. Key simulation
parameters are summarized in Table \ref{tab:Key-simulation-parameters}.
The simulation was run over 1000 subframes. We compare the performance
of the proposed algorithm with the following 3 baselines.
\begin{itemize}
\item \textbf{Baseline 1 (FFR with static synchronized ABS)}: Proportional
fair scheduling and static synchronized ABS are used. The ABS is transmitted
synchronously among the macro BS with $1/8$ blanking rate. Fractional
frequency reuse \cite{ghaffar2010fractional} with factor $1/3$ is
applied to the outer zone of each macro to protect the cell edge users. 
\item \textbf{Baseline 2 (FFR with dynamic synchronous ABS)} \cite{Pang:2012kx}:
Baseline 2 is the same as baseline 1, except that the ABS blanking
rate is dynamically chosen to maximize the proportional fairness utility.
\item \textbf{Baseline 3 (Clustered CoMP)}: 3 neighbor macro BSs and the
associated pico BSs form a cluster for cooperative zero-forcing (ZF)
\cite{somekh2009cooperative} with per BS power constraint. 
\end{itemize}

\begin{table}
\begin{centering}
\begin{tabular}{|l|l|}
\hline 
\textbf{Parameters} & \textbf{Values}\tabularnewline
\hline 
\hline 
Network layout & 19 eNBs, 3-cell sites,\tabularnewline
\multicolumn{1}{|l||}{} & 4 picos per sector (site)\tabularnewline
\hline 
Number of UE per cell site & 30\tabularnewline
\hline 
BS transmit power & Macro: 46 dBm, Pico 30 dBm\tabularnewline
\hline 
Channel model & IMT-Advanced Channel\tabularnewline
 & Model \cite[Annex B]{3gpp_Rel10}\tabularnewline
\hline 
Scheduling & Proportional Fair\tabularnewline
\hline 
Thermal noise & - 174 dBm/Hz\tabularnewline
\hline 
UE speed & 6 km/h\tabularnewline
\hline 
Bandwidth & 10 MHz\tabularnewline
\hline 
Number of subbands & 55\tabularnewline
\hline 
Cell selection bias & 9 dB\tabularnewline
\hline 
\end{tabular}
\par\end{centering}

\caption{\label{tab:Key-simulation-parameters}Key simulation parameters. }
\end{table}

\subsubsection{Throughput Evaluations}

\begin{figure}
\begin{centering}
\includegraphics[width=100mm]{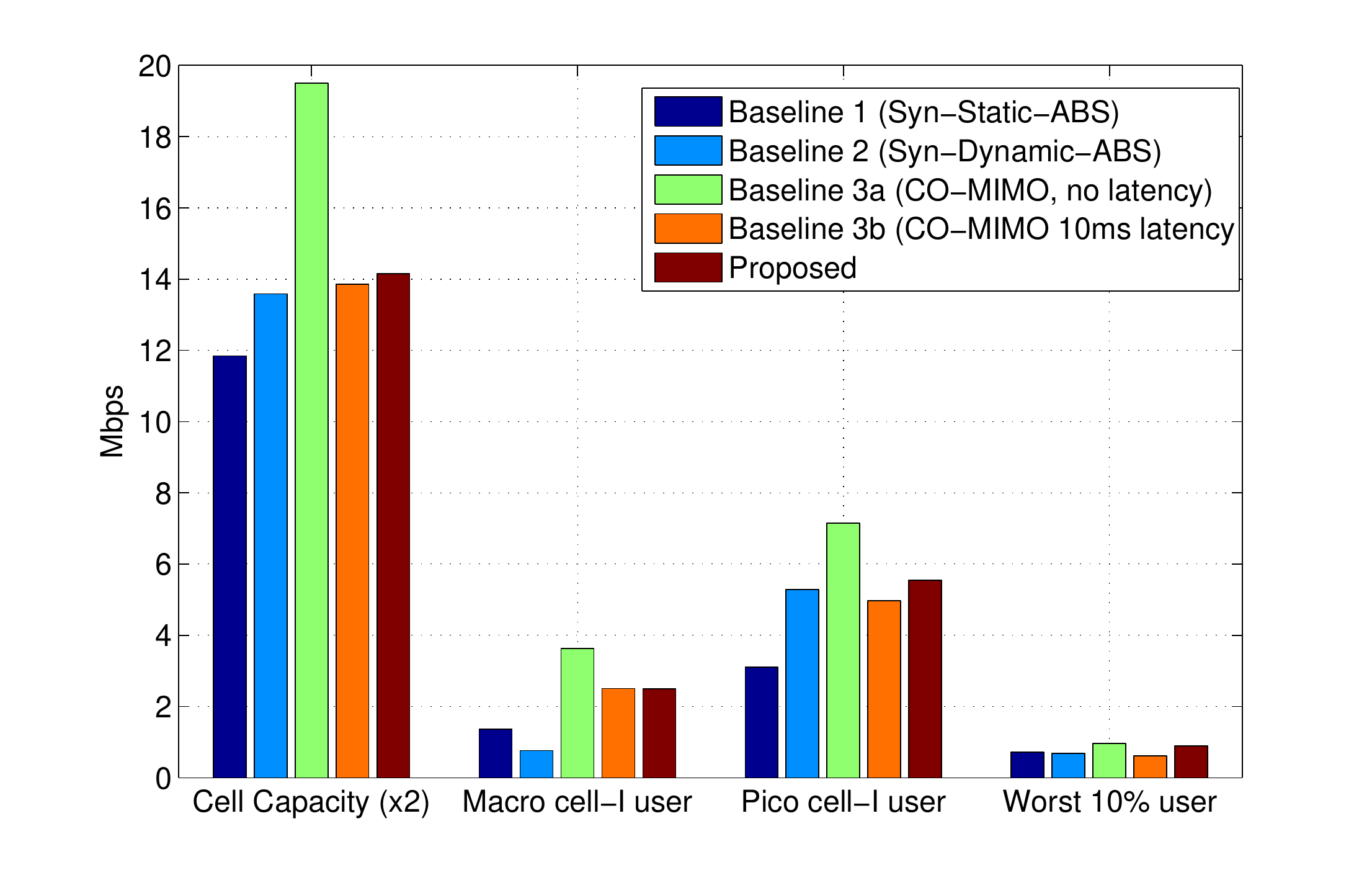}
\par\end{centering}

\caption{\label{fig:bar_throughput}{\small Throughput comparisons over different
RRM schemes and in different backhaul latency scenarios. The proposed
algorithm outperforms both baseline 1 and 2. It also outperforms baseline
3 when there is a backhaul latency.}}
\end{figure}

Fig. \ref{fig:bar_throughput} compares the throughput of different
RRM schemes. The proposed scheme outperforms baseline 1 and 2 over
all performance metrics. It also outperforms baseline 3 when $10$
ms backhaul latency for the signaling among BSs is considered. These
results demonstrate the superior performance and the robustness of
the proposed hierarchical RRM scheme w.r.t. signaling latency in backhaul.

Table \ref{tab:throughput-asymmetry} summaries the throughput performance
of baseline 2 and the proposed scheme under asymmetric network topologies,
where in each macro cell, the number of pico BS and the number of
users are Poisson distributed with mean $\lambda_{p}=4$ and $\lambda_{u}=30$,
respectively. The proposed scheme still outperforms baseline 2. In
particular, it enjoys 27\% throughput gain for the worst 10\% users.
As a comparison, the corresponding throughput gain for the worst 10\%
users is $6\%$ in the symmetric topology in Fig. \ref{fig:bar_throughput}.
This demonstrates that the proposed scheme can better adapt to dynamic
network loading. 
\begin{rem}
In the simulations, we have fixed the cell selection bias $\beta$
to be 9dB. For larger $\beta$, there will be more pico cell I-users
and more severe cross-tier interference from macro BS to pico cell
I-users. In this case, it is more critical to use better and more
fine-grained eICIC schemes to control the cross-tier interference.
As a result, the performance gap between the optimization based ABRB
control and those heuristic ABS controls becomes larger as increases.
\end{rem}
\begin{table}
\begin{centering}
\begin{tabular}{|c|c|c|c|}
\hline 
 & Baseline 2 & Proposed & Gain\tabularnewline
\hline 
Average cell capacity (Mbps) & 27.7 & 28.8 & 4\%\tabularnewline
\hline 
Macro cell I-users (Kbps) & 708 & 2351 & 232\%\tabularnewline
\hline 
Pico cell I-users (Kbps) & 5498 & 5548 & 1\%\tabularnewline
\hline 
worst 10\% users (Kbps) & 715 & 908 & 27\%\tabularnewline
\hline 
\end{tabular}
\par\end{centering}

\caption{\label{tab:throughput-asymmetry}{\small Performance evaluations on
a HetNet with asymmetric network topology, where in each macro cell,
the number of pico BS and the number of users are Poisson distributed
with mean $\lambda_{p}=4$ and $\lambda_{u}=30$, respectively.}}
\end{table}

\subsubsection{Convergence of the Proposed Hierarchical RRM Algorithm}

\begin{figure}
\begin{centering}
\includegraphics[width=85mm]{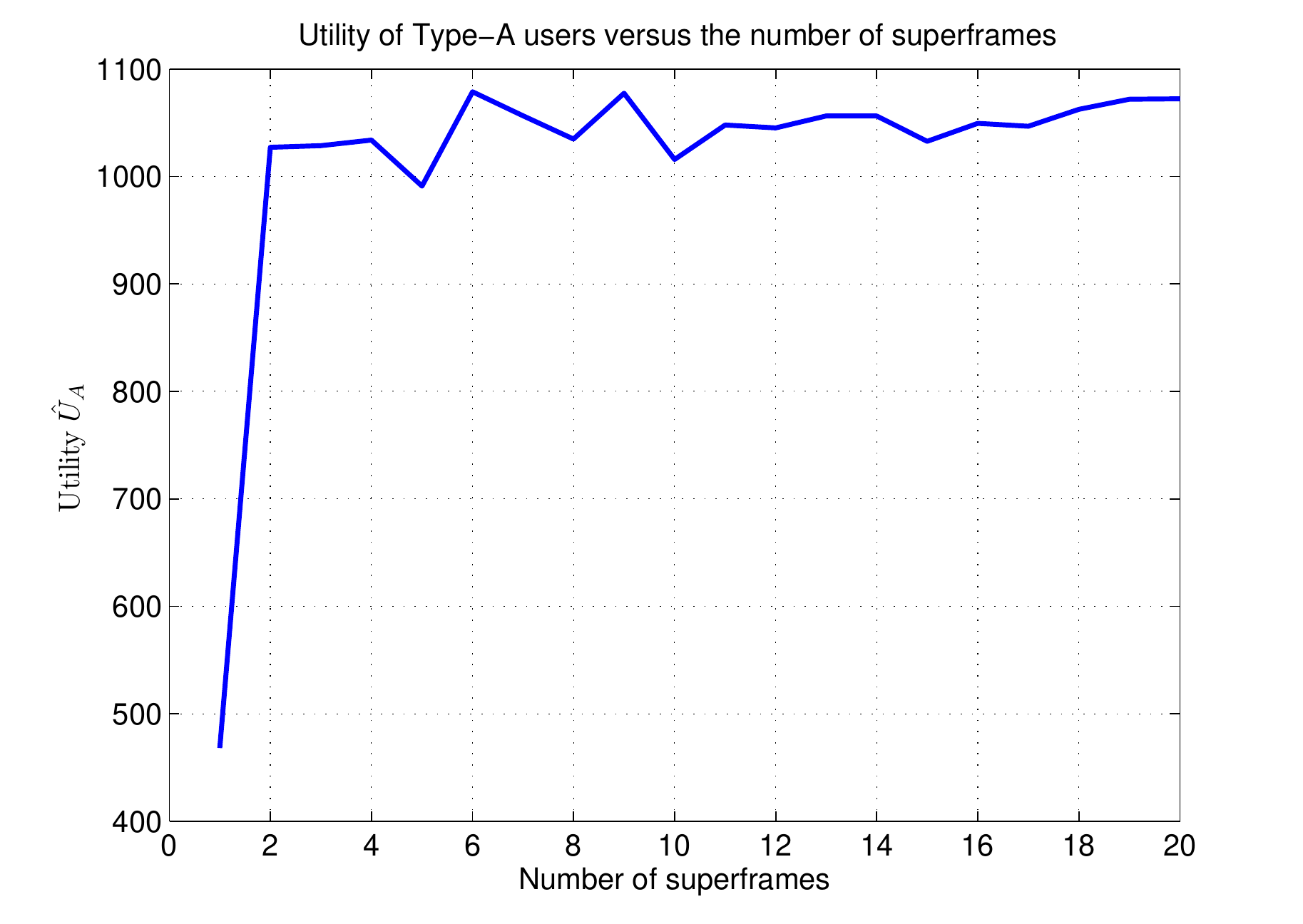}
\par\end{centering}

\caption{\label{fig:convA}{\small Utility of the Type A users versus the number
of super-frames}}
\end{figure}

Fig. \ref{fig:convA} shows the utility $\hat{U}_{A}$ in (\ref{eq:PLAre})
of the Type A users versus the number of super-frames. The utility
increases rapidly and then approaches to a steady state after only
$2-3$ updates. The figure demonstrates a fast convergence behavior
of Algorithm AO\_A. Similar convergence behavior was also observed
for Algorithm AO\_B and the simulation result is not shown here due
to limited space.

\subsection{Complexity }

We compare the complexity of the baselines and proposed RRM algorithms.
The complexity can be evaluated in the following 3 aspects. 

1) For the short term user scheduling, the proposed scheme and baseline
1-3 have the same complexity order of $\mathcal{O}\left(MK\right)$,
while the baseline 4 has a complexity of $\mathcal{O}\left(C_{4}MK\right)$
\cite{somekh2009cooperative}, where $C_{4}$ is a proportionality
constant that corresponds to some matrix and vector operations with
dimension $B_{\textrm{c}}$, and $B_{\textrm{c}}=15$ is the number
of BSs in each cooperative cluster. 

2) For the long term ABRB control variables $\mathbf{q}_{A}$ and
$\mathbf{q}_{B}$, as they are updated by solving standard convex
optimization problems in step 2 of Algorithm AO\_A and AO\_B respectively,
the complexities are polynomial w.r.t. the number of the associated
optimization variables. Specifically, for control variable $\mathbf{q}_{A}$,
the complexity is polynomial w.r.t. the number of macro BSs $N_{0}$.
For control variable $q_{B}$, the complexity is polynomial w.r.t.
the size of the ABRB profile $\mathcal{A}^{B*}$: $\left|\mathcal{A}^{B*}\right|\leq\left|\mathcal{U}_{B}\right|$.
In addition, they are only updated once in each super-frame. 

3) The ABRB profile $\mathcal{A}^{B*}$ is computed using Algorithm
B2 in every several super-frames to adapt to the large scale fading.
In step 1 of Algorithm B2, the complexity of solving the convex problem
(\ref{eq:fixthetaPdet}) is polynomial w.r.t. $\left|\Theta^{(i)}\right|\leq\left|\mathcal{U}_{B}\right|+1$.
In step 2 of Algorithm B2, if the MWIS algorithm in \cite{Brendel_NIPS10_MWIS}
is used to solve problem (\ref{eq:MWISP}), the complexity is $\mathcal{O}\left(|\mathcal{E}_{I}|\right)$,
where $|\mathcal{E}_{I}|$ is the number of edges in the interference
graph $\mathcal{G}_{I}\left(\mathcal{G}_{T}\right)=\left\{ \mathcal{L},\mathcal{E}_{I}\right\} $.

\section{Conclusion}

We propose a two-timescale hierarchical RRM for HetNet with dynamic
ABRB. To facilitate structural ABRB design for cross-tier and co-tier
interference, the $M$ subbands are partitioned into Type A and Type
B subbands. Consequently, the two timescale RRM problem is decomposed
into subproblems $\mathcal{P}_{A}$ and $\mathcal{P}_{B}$ which respectively
optimizes the ABRB control and user scheduling for the Type A and
Type B subbands. Both subproblems involve non-trivial multi-stage
optimization with exponential large solution space w.r.t. the number
of macro BSs $N_{0}$. We exploit the sparsity in the HetNet interference
graph and derive the structural properties to reduce the solution
space. Based on that, we propose two timescale AO algorithm to solve
$\mathcal{P}_{A}$ and $\mathcal{P}_{B}$. The overall solution is
asymptotically optimal at high SNR and has low complexity, low signaling
overhead as well as robust w.r.t. latency of backhaul signaling. 

\appendix

\subsection{Proof of Lemma \ref{lem:Optimality-of-Symmetric} \label{Proof-of-Lemma symP}}

Define the average rate region as
\[
\mathcal{R}\triangleq\underset{\Lambda}{\bigcup}\left\{ \mathbf{x}=\left[x_{k}\right]\in\mathbb{R}_{+}^{K}:\: x_{k}\le\overline{r}_{k}\left(\Lambda\right),\forall k\right\} .
\]
For any utility function that is concave and increasing w.r.t. to
the average data rates $\overline{\mathbf{r}}$, the optimal policy
of $\widetilde{\mathcal{P}}\left(\mathcal{G}_{T}\right)$ must achieve
a Pareto boundary point of $\mathcal{R}$. Hence, we only need to
show that any Pareto boundary point of $\mathcal{R}$ can be achieved
by a symmetric policy. Define the average rate region under fixed
$q_{s}$ as
\begin{eqnarray*}
\mathcal{R}\left(q_{s}\right) & \triangleq & \underset{\left\{ \left\{ Q_{m}\right\} ,\left\{ \pi_{m}\right\} \right\} }{\bigcup}\bigg\{\mathbf{x}=\left[x_{k}\right]\in\mathbb{R}_{+}^{K}:\\
 &  & x_{k}\le\overline{r}_{k}\left(\left\{ q_{s},\left\{ Q_{m}\right\} ,\left\{ \pi_{m}\right\} \right\} \right),\forall k\bigg\}.
\end{eqnarray*}
Then we only need to show that any Pareto boundary point of $\mathcal{R}\left(q_{s}\right)$
can be achieved by a symmetric policy $\Lambda^{s}=\left\{ q_{s},Q_{A},Q_{B},\pi_{A},\pi_{B}\right\} $.
Define the average mutual information region\emph{ }for subband $m\in\mathcal{M}_{A}$
as: 
\begin{eqnarray}
\mathcal{R}_{m} & \triangleq & \underset{\left\{ Q_{m},\pi_{m}\right\} }{\bigcup}\bigg\{\left[x_{k}\right]_{\forall k\in\mathcal{U}_{A}}\in\mathbb{R}_{+}^{\left|\mathcal{U}_{A}\right|}:\nonumber \\
 &  & x_{k}\le\overline{\mathcal{I}}_{m,k}\left(Q_{m},\pi_{m}\right),\forall k\in\mathcal{U}_{A}\bigg\}.\label{eq:Ratereg1G-1}
\end{eqnarray}
Define the average mutual information region\emph{ }for subband $m\in\mathcal{M}_{B}$
as: 
\begin{eqnarray}
\mathcal{R}_{m} & \triangleq & \underset{\left\{ Q_{m},\pi_{m}\right\} }{\bigcup}\bigg\{\left[x_{k}\right]_{\forall k\in\mathcal{U}_{B}}\in\mathbb{R}_{+}^{\left|\mathcal{U}_{B}\right|}:\nonumber \\
 &  & x_{k}\le\overline{\mathcal{I}}_{m,k}\left(Q_{m},\pi_{m}\right),\forall k\in\mathcal{U}_{B}\bigg\}.\label{eq:Ratereg2G-1}
\end{eqnarray}
It can be verified that $\mathcal{R}_{m},\forall m\in\mathcal{M}_{A}$
is a convex region in $\mathbb{R}_{+}^{\left|\mathcal{U}_{A}\right|}$
and $\mathcal{R}_{m},\forall m\in\mathcal{M}_{B}$ is a convex region
in $\mathbb{R}_{+}^{\left|\mathcal{U}_{B}\right|}$. Moreover, due
to the statistical symmetry of the subbands, we have 
\begin{eqnarray}
\mathcal{R}_{m} & = & \mathcal{R}_{m^{'}},\forall m,m^{'}\in\mathcal{M}_{A},\label{eq:Ra}\\
\mathcal{R}_{m} & = & \mathcal{R}_{m^{'}},\forall m,m^{'}\in\mathcal{M}_{B}.\label{eq:Rb}
\end{eqnarray}
Let $\mathcal{R}_{A}=\mathcal{R}_{m},\forall m\in\mathcal{M}_{A}$
and $\mathcal{R}_{B}=\mathcal{R}_{m},\forall m\in\mathcal{M}_{B}$.
From the convexity of $\mathcal{R}_{m},\forall m$ and (\ref{eq:Ra}-\ref{eq:Rb}),
we have
\begin{equation}
\mathcal{R}\left(q_{s}\right)=\left|\mathcal{M}_{A}\right|\mathcal{R}_{A}\times\left|\mathcal{M}_{B}\right|\mathcal{R}_{B}.\label{eq:Rqs}
\end{equation}
Hence, for any Pareto boundary point $\overline{\mathbf{r}}^{*}=\left[\overline{r}_{1}^{*},...,\overline{r}_{K}^{*}\right]^{T}$
of $\mathcal{R}\left(q_{s}\right)$, $\frac{1}{\left|\mathcal{M}_{A}\right|}\overline{\mathbf{r}}_{A}^{*}\triangleq\left[\frac{\overline{r}_{k}^{*}}{\left|\mathcal{M}_{A}\right|}\right]_{k\in\mathcal{U}_{A}}\in\mathbb{R}_{+}^{\left|\mathcal{U}_{A}\right|}$
is a Pareto boundary point of $\mathcal{R}_{A}$ and $\frac{1}{\left|\mathcal{M}_{B}\right|}\overline{\mathbf{r}}_{B}^{*}\triangleq\left[\frac{\overline{r}_{k}^{*}}{\left|\mathcal{M}_{B}\right|}\right]_{k\in\mathcal{U}_{B}}\in\mathbb{R}_{+}^{\left|\mathcal{U}_{B}\right|}$
is a Pareto boundary point of $\mathcal{R}_{B}$. Due to the statistical
symmetry of the subbands, there exists an ABRB control policy and
a user scheduling policy $\left\{ Q_{A}^{*},\pi_{A}^{*}\right\} $
such that $\frac{1}{\left|\mathcal{M}_{A}\right|}\overline{\mathbf{r}}_{A}^{*}$
can be achieved for all subbands $m\in\mathcal{M}_{A}$. Similarly,
there exists an ABRB control policy and a user scheduling policy $\left\{ Q_{B}^{*},\pi_{B}^{*}\right\} $
such that $\frac{1}{\left|\mathcal{M}_{B}\right|}\overline{\mathbf{r}}_{B}^{*}$
can be achieved for all subbands $m\in\mathcal{M}_{B}$. Hence, $\overline{\mathbf{r}}^{*}$
can be achieved using the symmetric policy $\Lambda^{s*}=\left\{ q_{s},Q_{A}^{*},Q_{B}^{*},\pi_{A}^{*},\pi_{B}^{*}\right\} $.
This completes the proof.

\subsection{Structural Properties of $\mathcal{P}_{A}$ for General Cases\label{sub:SPAgeneral}}

The formal definition of $\mathcal{A}_{n}$ and $\overline{\mathcal{A}}_{n}$
is 
\begin{eqnarray}
\mathcal{A}_{n} & = & \begin{cases}
\left\{ \mathbf{a}:\:\mathbf{a}\in\mathcal{A};\:\textrm{and}\: a_{n}=1\right\} , & \forall n\leq N_{0},\\
\left\{ \mathbf{a}:\:\mathbf{a}\in\mathcal{A};\:\textrm{and}\: a_{n^{'}}=0,\forall n^{'}\in\mathcal{B}_{n}\right\} , & \forall n>N_{0},
\end{cases}\nonumber \\
\overline{\mathcal{A}}_{n} & = & \mathcal{A}/\mathcal{A}_{n},\:\forall n\in\left\{ 1,...,N\right\} .\label{eq:AkAbk}
\end{eqnarray}
The result in Observation \ref{clm:Effect-of-ABRB-MI} is formally
stated in the following lemma.
\begin{lem}
\label{lem:sysmAn}For given CSI $\mathbf{H}_{m_{A}}\in\mathcal{H}$,
BS index $n\in\left\{ 1,...,N\right\} $ and user index $k\in\mathcal{U}_{n}\cap\mathcal{U}_{A}$,
the following are true:
\begin{enumerate}
\item $\forall\mathbf{a},\mathbf{a}^{'}\in\mathcal{A}_{n}$, we have $\Gamma_{m_{A}}^{n}\left(\mathbf{a}\right)=\Gamma_{m_{A}}^{n}\left(\mathbf{a}^{'}\right)$
and 
\[
\mathcal{I}_{m_{A},k}\left(\mathbf{a},\mathbf{H}_{m_{A}},\rho_{m_{A}}^{n}\right)=\mathcal{I}_{m_{A},k}\left(\mathbf{a}^{'},\mathbf{H}_{m_{A}},\rho_{m_{A}}^{n}\right),
\]
$\forall\rho_{m_{A}}^{n}\in\Gamma_{m_{A}}^{n}\left(\mathbf{a}\right)$.
The same is true if we replace $\mathcal{A}_{n}$ with $\overline{\mathcal{A}}_{n}$.
\item For given ABRB patterns $\mathbf{a}\in\mathcal{A}_{n},\mathbf{a}^{'}\in\overline{\mathcal{A}}_{n}$
and user scheduling vector $\rho_{m_{A}}^{n'}\in\Gamma_{m_{A}}^{n}\left(\mathbf{a}^{'}\right)$,
there exists $\rho_{m_{A}}^{n}\in\Gamma_{m_{A}}^{n}\left(\mathbf{a}\right)$
such that $\mathcal{I}_{m_{A},k}\left(\mathbf{a},\mathbf{H}_{m_{A}},\rho_{m_{A}}^{n}\right)\geq\mathcal{I}_{m_{A},k}\left(\mathbf{a}^{'},\mathbf{H}_{1},\rho_{m_{A}}^{n'}\right)$.
\end{enumerate}
\end{lem}
\begin{IEEEproof}
By Definition \ref{def:INuser}, there is no inter-cell interference
for macro cell N-users. By the definition of $\Gamma_{m_{A}}^{n}\left(\mathbf{a}\right)$,
a pico cell I-user $k\in\mathcal{U}_{n}\cap\mathcal{U}_{A}$ cannot
be scheduled for transmission if any of the neighbor macro BSs in
$\mathcal{B}_{n}$ is transmitting data subframe (i.e., the current
ABRB pattern $\mathbf{a}\in\overline{\mathcal{A}}_{n}$). On the other
hand, if all of the neighbor macro BSs is transmitting ABRB (i.e.,
the current ABRB pattern $\mathbf{a}\in\mathcal{A}_{n}$), the interference
from the macro BSs is negligible. Finally, by Definition \ref{def:INuser},
there is no inter-cell interference for pico cell N-users. Then Lemma
1 follows straightforwardly from the above analysis and the definition
of $\mathcal{A}_{n},\overline{\mathcal{A}}_{n}$.
\end{IEEEproof}

\begin{rem}
\label{rem:usefulAssumptions}Note that in the proof of Lemma \ref{lem:sysmAn},
we have used the fact that the inter-cell interference seen at a N-user
is negligible. We have also used Assumption \ref{asm:picoIedge},
which states that the sets of neighbor macro BSs of the pico cell
I-users belonging to the same pico cell are identical.
\end{rem}

For general cases, the result in Observation \ref{clm:QAreduction}
is stated in the following theorem.
\begin{thm}
[Policy Space Reduction for $Q_{A}$]\label{thm:ABSreduce}Given
a marginal probability vector that each macro BS is transmitting ABRB
$\mathbf{q}_{A}=\left\{ q_{j}^{A},\: j=1,...,N_{0}\right\} $, the
optimal ABRB control policy of $\mathcal{P}_{A}\left(\mathcal{G}_{T}\right)$
conditioned on \textup{$\mathbf{q}_{A}$, denoted by $Q_{A|\mathbf{q}_{A}}^{*}$,}
has the following synchronous ABRB structure: 

(a) Let $\varsigma$ be a permutation such that $q_{\varsigma(1)}^{A}\leq q_{\varsigma(2)}^{A},...,\leq q_{\varsigma(N_{0})}^{A}$.
The support of \textup{$Q_{A|\mathbf{q}_{A}}^{*}$} has only $N_{0}+1$
active ABRB patterns $\mathcal{A}^{A}\left(\mathbf{q}_{A}\right)=\left\{ \mathbf{a}^{A}(1),...,\mathbf{a}^{A}(N_{0}+1)\right\} $,
where $\mathbf{a}^{A}(j)=\left[a_{1}^{A}(j),...,a_{N_{0}}^{A}(j)\right],j=1,...,N_{0}+1$
with $a_{\varsigma(i)}^{A}(j)=1,\:1\leq i\leq N_{0}+1-j$ and $a_{\varsigma(i)}^{A}(j)=0,\: N_{0}+1-j<i\leq N_{0}$.

(b) Define $q_{0}^{A}=0$, $q_{N_{0}+1}^{A}=1$, $\varsigma(0)=0$,
and $\varsigma(N_{0}+1)=N_{0}+1$. Then $Q_{A|\mathbf{q}_{A}}^{*}\left(\mathbf{a}^{A}(j)\right)=q_{\varsigma(j)}^{A}-q_{\varsigma(j-1)}^{A},\: j=1,...,N_{0}+1$,
and $Q_{A|\mathbf{q}_{A}}^{*}\left(\mathbf{a}\right)=0,\:\forall\mathbf{a}\notin\mathcal{A}^{A}\left(\mathbf{q}_{A}\right)$.\end{thm}
\begin{IEEEproof}
By Lemma \ref{lem:sysmAn}, for given marginal probabilities $\mathbf{q}_{A}$,
the average mutual information region will be maximized if we maximize
$\sum_{\mathbf{a}\in\mathcal{A}_{n}}Q_{A}\left(\mathbf{a}\right)$
for all $1\leq n\leq N$. For $n\leq N_{0}$, we have $\sum_{\mathbf{a}\in\mathcal{A}_{n}}Q_{A}\left(\mathbf{a}\right)=q_{n}^{A}$.
For $n>N_{0}$, we have $\sum_{\mathbf{a}\in\mathcal{A}_{n}}Q_{A}\left(\mathbf{a}\right)\leq\underset{j\in\mathcal{B}_{n}}{\textrm{min}}\left(q_{j}^{A}\right)$,
where the equality holds if and only if $Q_{A}$ has the structure
in Theorem \ref{thm:ABSreduce}. This completes the proof.
\end{IEEEproof}

An example of synchronous ABRB is illustrated in Fig. \ref{fig:ABS_reduction}.
The following theorem is a general version of Observation \ref{clm:Policy-Space-ReductionpiA}.
\begin{thm}
[Policy Space Reduction for $\pi_{A}$]\label{thm:Policy-Space-ReductionpiA}There
exists optimal user scheduling policy $\pi_{A}^{*}$ for $\mathcal{P}_{A}\left(\mathcal{G}_{T}\right)$
such that $\pi_{A}^{*}\in\Xi_{A}^{*}$, where $\Xi_{A}^{*}\triangleq\left\{ \pi_{A}=\left\{ \pi_{A}^{1},...,\pi_{A}^{N}\right\} :\:\pi_{A}^{n},\forall n,\:\textrm{satisfies}\:(\ref{eq:piAC1})\:\textrm{and}\:(\ref{eq:piAC2})\right\} $.
\begin{eqnarray}
\pi_{A}^{n}\left(\mathbf{a},\mathbf{H}_{m_{A}}\right)=\pi_{A}^{n}\left(\mathbf{a}^{'},\mathbf{H}_{m_{A}}\right),\forall\mathbf{a},\mathbf{a}^{'}\in\mathcal{A}_{n},\mathbf{H}_{m_{A}}\in\mathcal{H}.\label{eq:piAC1}\\
\pi_{A}^{n}\left(\mathbf{a},\mathbf{H}_{m_{A}}\right)=\pi_{A}^{n}\left(\mathbf{a}^{'},\mathbf{H}_{m_{A}}\right)\forall\mathbf{a},\mathbf{a}^{'}\in\overline{\mathcal{A}}_{n},\mathbf{H}_{m_{A}}\in\mathcal{H}.\label{eq:piAC2}
\end{eqnarray}
\end{thm}
\begin{IEEEproof}
Define the\emph{ achievable mutual information region }for subband
$m_{A}$ as 
\begin{equation}
\mathcal{R}_{m_{A}}\triangleq\underset{Q_{A}\in\mathcal{Q},\pi_{A}}{\bigcup}\left\{ \left[x_{k}\right]_{\forall k\in\mathcal{U}_{A}}:x_{k}\le\overline{\mathcal{I}}_{m_{A},k}\left(Q_{A},\pi_{A}\right)\right\} ,\label{eq:Ratereg1G}
\end{equation}
where $\mathcal{Q}=\left\{ Q_{A}:\:\sum_{\mathbf{a}\in\mathcal{A}}Q_{A}\left(\mathbf{a}\right)=1;Q_{A}\left(\mathbf{a}\right)\geq0,\forall\mathbf{a}\in\mathcal{A}\right\} $.
It can be verified that $\mathcal{R}_{m_{A}}$ is a convex region
in $\mathbb{R}_{+}^{\left|\mathcal{U}_{A}\right|}$. Since the utility
function $U_{A}\left(Q_{A},\pi_{A}\right)$ is concave and increasing
w.r.t. $\overline{\mathcal{I}}_{m_{A},k}\left(Q_{A},\pi_{A}\right)$,
the optimal policy $Q_{A}^{*},\pi_{A}^{*}$ must achieve a Pareto
boundary point of $\mathcal{R}_{m_{A}}$. For given ABRB pattern $\mathbf{a}$
and BS $n$, define a region as
\[
\mathcal{R}_{m_{A}}^{n}\left(\mathbf{a}\right)\triangleq\underset{\pi_{A}}{\bigcup}\left\{ \left[x_{k}\right]_{\forall k\in\mathcal{U}_{A}\cap\mathcal{U}_{n}}:\: x_{k}\le I_{m_{A},k}\left(\pi_{A},\mathbf{a}\right)\right\} .
\]
It can be verified that $\mathcal{R}_{m_{A}}^{n}\left(\mathbf{a}\right)$
is a convex region. From Lemma \ref{lem:sysmAn}, we have 
\begin{eqnarray}
\mathcal{R}_{m_{A}}^{n}\left(\mathbf{a}\right) & = & \mathcal{R}_{m_{A}}^{n}\left(\mathbf{a}^{'}\right),\forall\mathbf{a},\mathbf{a}^{'}\in\mathcal{A}_{n}.\label{eq:Rma}\\
\mathcal{R}_{m_{A}}^{n}\left(\mathbf{a}\right) & = & \mathcal{R}_{m_{A}}^{n}\left(\mathbf{a}^{'}\right),\forall\mathbf{a},\mathbf{a}^{'}\in\overline{\mathcal{A}}_{n}.\label{eq:Rmabar}
\end{eqnarray}
For convenience, define 
\begin{eqnarray*}
e_{k}^{*} & \triangleq & \sum_{\mathbf{a}\in\mathcal{A}_{n}}Q_{A}^{*}\left(\mathbf{a}\right)I_{m_{A},k}\left(\pi_{A}^{*},\mathbf{a}\right)/\sum_{\mathbf{a}\in\mathcal{A}_{n}}Q_{A}^{*}\left(\mathbf{a}\right),\\
\overline{e}_{k}^{*} & \triangleq & \sum_{\mathbf{a}\in\overline{\mathcal{A}}_{n}}Q_{A}\left(\mathbf{a}\right)I_{m_{A},k}\left(\pi_{A}^{*},\mathbf{a}\right)/\sum_{\mathbf{a}\in\overline{\mathcal{A}}_{n}}Q_{A}^{*}\left(\mathbf{a}\right).
\end{eqnarray*}
Then $\forall n\in\left\{ 1,...,N\right\} ,k\in\mathcal{U}_{A}\cap\mathcal{U}_{n}$,
we have $\overline{\mathcal{I}}_{m_{A},k}\left(Q_{A}^{*},\pi_{A}^{*}\right)=\sum_{\mathbf{a}\in\mathcal{A}_{n}}Q_{A}^{*}\left(\mathbf{a}\right)e_{k}^{*}+\sum_{\mathbf{a}\in\overline{\mathcal{A}}_{n}}Q_{A}^{*}\left(\mathbf{a}\right)\overline{e}_{k}^{*}$.
From (\ref{eq:Rma}-\ref{eq:Rmabar}) and the fact that $\left[\overline{\mathcal{I}}_{m_{A},k}\left(Q_{A}^{*},\pi_{A}^{*}\right)\right]_{\forall k\in\mathcal{U}_{A}}$
is a Pareto boundary point of $\mathcal{R}_{m_{A}}$, it follows that
$\left[e_{k}^{*}\right]_{\forall k\in\mathcal{U}_{A}\cap\mathcal{U}_{n}}$
is a Pareto boundary point of $\mathcal{R}_{m_{A}}^{n}\left(\mathbf{a}\right),\forall\mathbf{a}\in\mathcal{A}_{n}$
and $\left[\overline{e}_{k}^{*}\right]_{\forall k\in\mathcal{U}_{A}\cap\mathcal{U}_{n}}$
is a Pareto boundary point of $\mathcal{R}_{m_{A}}^{n}\left(\mathbf{a}\right),\forall\mathbf{a}\in\overline{\mathcal{A}}_{n}$.
Hence, there exists user scheduling policy $\pi_{A}^{\circ}\in\Xi_{A}^{*}$
satisfying $I_{m_{A},k}\left(\pi_{A}^{\circ},\mathbf{a}\right)=e_{k}^{*},\forall\mathbf{a}\in\mathcal{A}_{n}$
and $I_{m_{A},k}\left(\pi_{A}^{\circ},\mathbf{a}\right)=\overline{e}_{k}^{*},\forall\mathbf{a}\in\overline{\mathcal{A}}_{n}$
for all $k\in\mathcal{U}_{A}\cap\mathcal{U}_{n}$. Then it follows
that $\left[\overline{\mathcal{I}}_{m_{A},k}\left(Q_{A}^{*},\pi_{A}^{*}\right)\right]_{\forall k\in\mathcal{U}_{A}}$
can be achieved by the control policy $Q_{A}^{*},\pi_{A}^{\circ}\in\Xi_{A}^{*}$.
\end{IEEEproof}

\subsection{Proof of Corollary \ref{cor:Equivalent-Problem-PA} \label{sub:Proof-of-CorollayPA}}

The first part of the corollary follows straightforward from Theorem
\ref{thm:ABSreduce} and \ref{clm:Policy-Space-ReductionpiA}. We
only need to prove that problem (\ref{eq:PLAre}) is bi-convex. The
average mutual information in (\ref{eq:UAobj}) can be expressed as{\small 
\begin{eqnarray}
 & \overline{\mathcal{I}}_{m_{A},k}\left(Q_{A|\mathbf{q}_{A}}^{*},\pi_{A}\right)=\sum_{\mathbf{a}\in\mathcal{A}_{b_{k}}}Q_{A|\mathbf{q}_{A}}^{*}\left(\mathbf{a}\right)I_{m_{A},k}\left(\pi_{A},\mathbf{a}\right)\nonumber \\
 & +\sum_{\mathbf{a}\in\overline{\mathcal{A}}_{b_{k}}}Q_{A|\mathbf{q}_{A}}^{*}\left(\mathbf{a}\right)I_{m_{A},k}\left(\pi_{A},\mathbf{a}\right)\nonumber \\
= & \begin{cases}
\left(1-q_{b_{k}}^{A}\right)e_{k}^{A}\left(\pi_{A}\right), & k\in\mathcal{U}_{\textrm{MN}}\\
\underset{j\in\mathcal{B}_{b_{k}}}{\textrm{min}}q_{j}^{A}\left(e_{k}^{A}\left(\pi_{A}\right)-\overline{e}_{k}^{A}\left(\pi_{A}\right)\right)+\overline{e}_{k}^{A}\left(\pi_{A}\right), & k\in\mathcal{U}_{\textrm{PN}}\\
\underset{j\in\mathcal{B}_{b_{k}}}{\textrm{min}}q_{j}^{A}e_{k}^{A}\left(\pi_{A}\right), & k\in\mathcal{U}_{\textrm{PI}}
\end{cases}\label{eq:IBARAk}
\end{eqnarray}
}where $e_{k}^{A}\left(\pi_{A}\right)=I_{m_{A},k}\left(\pi_{A},\mathbf{a}\right),\:\forall\mathbf{a}\in\mathcal{A}_{b_{k}}$,
$\overline{e}_{k}^{A}\left(\pi_{A}\right)=I_{m_{A},k}\left(\pi_{A},\mathbf{a}\right),\:\forall\mathbf{a}\in\overline{\mathcal{A}}_{b_{k}}$,
$\mathcal{U}_{\textrm{MN}}$ is the set of macro cell N-users, $\mathcal{U}_{\textrm{PN}}$
is the set of pico cell N-users, and $\mathcal{U}_{\textrm{PI}}$
is the set of pico cell I-users. It is easy to verify that $\overline{\mathcal{I}}_{m_{A},k}\left(Q_{A|\mathbf{q}_{A}}^{*},\pi_{A}\right),\forall k\in\mathcal{U}_{A}$
is a concave function w.r.t. $\mathbf{q}_{A}$ for fixed $\pi_{A}$.
Using the vector composition rule for concave function \cite{Boyd_04Book_Convex_optimization},
the objective in (\ref{eq:PLAre}) is also concave w.r.t. $\mathbf{q}_{A}$
and thus problem (\ref{eq:PLAre}) is convex w.r.t. $\mathbf{q}_{A}$
for fixed $\pi_{A}$. For fixed $\mathbf{q}_{A}$, $\overline{\mathcal{I}}_{m_{A},k}\left(Q_{A|\mathbf{q}_{A}}^{*},\pi_{A}\right),\forall k\in\mathcal{U}_{A}$
is a linear function of the user scheduling variables $\left\{ \pi_{A}^{n}\left(\mathbf{a}_{m_{A}},\mathbf{H}_{m_{A}}\right)\right\} $.
Hence problem (\ref{eq:PLAre}) is also convex w.r.t. $\pi_{A}$ for
fixed $\mathbf{q}_{A}$.

\subsection{Proof of Theorem \ref{thm:Global-Convergence-LA}\label{sub:Proof-of-TheoremLA}}

It is clear that $\hat{U}_{A}\left(\mathbf{q}_{A}^{(T+1)},\pi_{A}^{(T)}\right)\geq\hat{U}_{A}\left(\mathbf{q}_{A}^{(T)},\pi_{A}^{(T-1)}\right)$.
By the assumption in Theorem \ref{thm:Global-Convergence-LA}, we
have $\hat{U}_{A}\left(\mathbf{q}_{A}^{(T+1)},\pi_{A}^{(T)}\right)>\hat{U}_{A}\left(\mathbf{q}_{A}^{(T)},\pi_{A}^{(T-1)}\right)$
if $\mathbf{q}_{A}^{(T)},\pi_{A}^{(T-1)}$ is not a fixed point of
$\mathcal{F}$. Combining the above and the fact that $\hat{U}_{A}$
is upper bounded, AO\_A must converge to a fixed point $\left[\tilde{\mathbf{q}}_{A},\tilde{\pi}_{A}\right]\in\Delta$.
The rest is to prove that any $\left[\tilde{\mathbf{q}}_{A},\tilde{\pi}_{A}\right]\in\Delta$
is globally optimal for problem (\ref{eq:PLAre}).

Note that problem (\ref{eq:PLAre}) is equivalent to the problem 
\begin{equation}
\max\sum_{k\in\mathcal{U}_{A}}w_{k}u\left(\mathcal{I}_{k}\right),\:\textrm{s.t.}\:\left[\mathcal{I}_{k}\right]_{\forall k\in\mathcal{U}_{A}}\in\mathcal{R}_{m_{A}}.\label{eq:EquPLApre}
\end{equation}
Since the objective in (\ref{eq:EquPLApre}) is a concave function
w.r.t. $\mathcal{I}_{k},\:\forall k\in\mathcal{U}_{A}$, and $\mathcal{R}_{m_{A}}$
is a convex region, the following Lemma holds.
\begin{lem}
[Optimality Condition for (\ref{eq:PLAre})]\label{lem:Opt-Cond-PA}A
solution $q_{A}^{*}$, $\pi_{A}^{*}$ is optimal for problem (\ref{eq:PLAre})
if and only if its average mutual information $\overline{\mathcal{I}}_{m_{A},k}^{*}\triangleq\overline{\mathcal{I}}_{m_{A},k}\left(Q_{A|\mathbf{q}_{A}^{*}}^{*},\pi_{A}^{*}\right),\: k\in\mathcal{U}_{A}$
satisfies{\small 
\[
\sum_{k\in\mathcal{U}_{A}}w_{k}\frac{\partial u\left(r\right)}{\partial r}|_{r=\overline{\mathcal{I}}_{m_{A},k}^{*}}\left(\overline{\mathcal{I}}_{m_{A},k}^{*}-\mathcal{I}_{k}\right)\geq0,\forall\left[\mathcal{I}_{k}\right]_{\forall k\in\mathcal{U}_{A}}\in\mathcal{R}_{m_{A}}
\]
}{\small \par}
\end{lem}

According to the step 1 of AO\_A, for any ABRB pattern $\mathbf{a}_{m_{A}}\in\mathcal{A}$
and CSI realization $\mathbf{H}_{m_{A}}\in\mathcal{H}$, the user
scheduling vector $\tilde{\rho}_{m_{A}}^{n}$ of BS $n$ under $\tilde{\pi}_{A}$
is the optimal solution of
\begin{equation}
\underset{\rho_{m_{A}}^{n}\in\Gamma_{m_{A}}^{n}\left(\mathbf{a}_{m_{A}}\right)}{\textrm{max}}\sum_{k\in\mathcal{U}_{A}\cap\mathcal{U}_{n}}\tilde{\mu}_{k}\mathcal{I}_{m_{A},k}\left(\mathbf{a}_{m_{A}},\mathbf{H}_{m_{A}},\rho_{m_{A}}^{n}\right),\label{eq:optpiAtuta}
\end{equation}
where $\tilde{\mu}_{k}=w_{k}\frac{\partial u\left(r\right)}{\partial r}|_{r=\tilde{\overline{\mathcal{I}}}_{m_{A},k}}$
and $\tilde{\overline{\mathcal{I}}}_{m_{A},k}\triangleq\overline{\mathcal{I}}_{m_{A},k}\left(Q_{A|\tilde{\mathbf{q}}_{A}}^{*},\tilde{\pi}_{A}\right)$
is the average mutual information under $\tilde{\mathbf{q}}_{A},\tilde{\pi}_{A}$.
Combining (\ref{eq:optpiAtuta}) and the fact that $\tilde{\mathbf{q}}_{A}$
is the optimal solution of problem (\ref{eq:PLAre}) with fixed $\tilde{\pi}_{A}$,
we have $\sum_{k\in\mathcal{U}_{A}}\tilde{\mu}_{k}\tilde{\overline{\mathcal{I}}}_{m_{A},k}\geq\sum_{k\in\mathcal{U}_{A}}\tilde{\mu}_{k}\overline{\mathcal{I}}_{m_{A},k}\left(Q_{A|\mathbf{q}_{A}}^{*},\tilde{\pi}_{A}\right)\geq\sum_{k\in\mathcal{U}_{A}}\tilde{\mu}_{k}\overline{\mathcal{I}}_{m_{A},k}\left(Q_{A|\mathbf{q}_{A}}^{*},\pi_{A}\right),\:\forall\mathbf{q}_{A}\in\mathcal{Q}^{A},\forall\pi_{A}$.
This implies that $\tilde{\mathbf{q}}_{A},\tilde{\pi}_{A}$ satisfy
the optimality condition in Lemma \ref{lem:Opt-Cond-PA}, and thus
is the globally optimal solution.

\subsection{Optimization of ABRB Profile \label{sub:Proof-of-TheoremEquPB}}

For any MIS $\mathcal{V}$, define $\vec{\mathcal{I}}\left(\mathcal{V}\right)\triangleq\left[\mathcal{I}_{k}\left(\mathcal{V}\right)\right]_{\forall k\in\mathcal{U}_{B}}$,
where 
\[
\mathcal{I}_{k}\left(\mathcal{V}\right)\triangleq\begin{cases}
\textrm{log}\left(1+P_{n}\sigma_{k,n}^{2}\right), & \textrm{if}\: l_{k}\in\mathcal{V},\\
0, & \textrm{otherwise},
\end{cases}
\]
The ABRB profile optimization algorithm is given below.

\smallskip{}

\textit{Algorithm B2} (Algorithm for solving $\mathcal{P}_{B}^{II}\left(\mathcal{G}_{T}\right)$): 

\textbf{\small Initialization}{\small : Find initial $\Theta^{(0)}=\left\{ \mathcal{V}_{1}^{(0)},...,\mathcal{V}_{\left|\Theta^{(0)}\right|}^{(0)}\right\} \subseteq\Theta_{T}\left(\mathcal{G}_{T}\right)$
such that $\mathcal{V}_{1}^{(0)}\cup\mathcal{V}_{2}^{(0)}...\cup\mathcal{V}_{\left|\Theta^{(0)}\right|}^{(0)}=\mathcal{L}$.
Set $i=0$.}{\small \par}

\textbf{\small Step 1}{\small{} (Update the coefficients $\check{\mathbf{q}}$):
For fixed $\Theta^{(i)}$, obtain the optimal solution $\check{\mathbf{q}}^{(i+1)}=\left[\check{q}_{j}^{(i+1)}\right]_{j=1,...,\left|\Theta^{(i)}\right|}$
of the following convex optimization problem
\begin{eqnarray}
 & \check{U}_{B}\left(\Theta^{(i)}\right)\triangleq\underset{\check{\mathbf{q}}}{\max}\sum_{k\in\mathcal{U}_{B}}w_{k}u\left(\sum_{j=1}^{\left|\Theta^{(i)}\right|}\check{q}_{j}\mathcal{I}_{k}\left(\mathcal{V}_{j}^{(i)}\right)\right),\label{eq:fixthetaPdet}\\
 & \textrm{s.t.}\:\check{q}_{j}\in\left[0,1\right],\forall j\:\textrm{and}\:\sum_{j=1}^{\left|\Theta^{(i)}\right|}\check{q}_{j}=1,\nonumber 
\end{eqnarray}
where $\check{\mathbf{q}}=\left[\check{q}_{j}\right]_{j=1,...,\left|\Theta^{(i)}\right|}$,
and $\mathcal{V}_{j}^{(i)}$ is the $j^{\textrm{th}}$ MIS in $\Theta^{(i)}$. }{\small \par}

\textbf{\small Step 2 }{\small (Update the set of MISs $\Theta$):}\textbf{\small{}
}{\small Let}\textbf{\small{} $\Theta^{(i+1)}=\left\{ \mathcal{V}_{j}^{(i)}:\:\check{q}_{j}^{(i+1)}>0\right\} \cup\mathcal{V}^{(i+1)}$}{\small ,
where $\mathcal{V}^{(i+1)}$ is given by}\textbf{\small{} 
\begin{equation}
\mathcal{V}^{(i+1)}=\underset{\mathcal{V}\in\Theta_{T}\left(\mathcal{G}_{T}\right)}{\textrm{argmax}}\:\sum_{k\in\mathcal{U}_{B}}\check{\mu}_{k}^{(i)}\vec{\mathcal{I}}_{k}\left(\mathcal{V}\right),\label{eq:MWISP}
\end{equation}
}{\small where}\textbf{\small{} $\check{\mu}_{k}^{(i)}=w_{k}\frac{\partial u\left(r\right)}{\partial r}|_{r=\sum_{j=1}^{\left|\Theta^{(i)}\right|}\check{q}_{j}^{(i+1)}\mathcal{I}_{k}\left(\mathcal{V}_{j}^{(i)}\right)}$}{\small .}{\small \par}

{\small If}\textbf{\small{} $\left|\check{U}_{B}\left(\Theta^{(i)}\right)-\check{U}_{B}\left(\Theta^{(i-1)}\right)\right|>\varepsilon$}{\small ,}\textbf{\small{}
}{\small let}\textbf{\small{} $i=i+1$ }{\small and return to Step
1. Otherwise, terminate the algorithm with $\Theta^{*}\triangleq\left\{ \mathcal{V}_{1}^{*},...,\mathcal{V}_{\left|\Theta^{*}\right|}^{*}\right\} =\left\{ \mathcal{V}_{j}^{(i)}:\:\check{q}_{j}^{(i+1)}>0\right\} $
and $\mathcal{A}^{B*}=\left\{ \mathbf{a}^{B*}(j),\: j=1,...,\left|\Theta^{*}\right|\right\} $,
where $\mathbf{a}^{B*}(j)=\left[a_{1}^{B*}(j),...,a_{N_{0}}^{B*}(j)\right]$
with $a_{n}^{B*}(j)=1,\:\forall n\in\mathcal{N}\left(\mathcal{V}_{j}^{*}\right)$
and $a_{n}^{B*}(j)=0,\:\forall n\notin\mathcal{N}\left(\mathcal{V}_{j}^{*}\right)$.}{\small \par}

\smallskip{}

The convergence and asymptotic optimality of Algorithm B2 is proved
in the following theorem.
\begin{thm}
[Asymptotically Optimal ABRB Profile]\label{thm:Asymptotic-equivalence-ofPB}Algorithm
B2 always converges to an ABRB profile $\mathcal{A}^{B*}$ with $\left|\mathcal{A}^{B*}\right|\leq\left|\mathcal{U}_{B}\right|$.
Furthermore, the converged result \textup{$\mathcal{A}^{B*}$} is
asymptotically optimal for high SNR. i.e. $\underset{P\rightarrow\infty}{\textrm{lim}}\left(U_{B}^{*}-U_{B}^{I}\left(\mathcal{A}^{B*}\right)\right)/U_{B}^{*}=0$,
where $P_{n}=\alpha_{n}P,\:\forall n=1,...,N_{0}$ for some positive
constants $\alpha_{n}$'s, and $U_{B}^{*}$ is the optimal objective
value of $\mathcal{P}_{B}\left(\mathcal{G}_{T}\right)$.\end{thm}
\begin{IEEEproof}
Consider problem $\check{\mathcal{P}}_{B}\left(\mathcal{G}_{T}\right)$
which is the same as $\mathcal{P}_{B}\left(\mathcal{G}_{T}\right)$
except that there are two differences: 1) the fading channel $\mathbf{H}\left(t\right)$
is replaced by a \textit{deterministic channel} with the channel gain
between BS $n$ and user $k$ given by the corresponding large scale
fading factor $\sigma_{k,n}^{2}$; 2) an additional constraint is
added to the user scheduling policy such that any two links $l_{k},l_{k^{'}}$
having an edge (i.e., $e\left(l_{k},l_{k^{'}}\right)=1$) in the interference
graph $\mathcal{G}_{I}\left(\mathcal{G}_{T}\right)$ cannot be scheduled
for transmission simultaneously. It can be shown that the optimal
solution of problem $\check{\mathcal{P}}_{B}\left(\mathcal{G}_{T}\right)$
is asymptotically optimal for $\mathcal{P}_{B}\left(\mathcal{G}_{T}\right)$
at high SNR. Moreover, using the fact that the achievable mutual information
region in the deterministic channel is a convex polytope with $\left\{ \vec{\mathcal{I}}\left(\mathcal{V}\right),\forall\mathcal{V}\in\Theta_{T}\left(\mathcal{G}_{T}\right)\right\} $
as the set of\textit{ Pareto boundary} vertices, it can be shown that
$\check{\mathcal{P}}_{B}\left(\mathcal{G}_{T}\right)$ is equivalent
to the following problem 
\begin{eqnarray}
\underset{\Theta}{\max} & \left\{ \underset{\check{\mathbf{q}}\in\check{\mathcal{Q}}\left(\Theta\right)}{\max}\sum_{k\in\mathcal{U}_{B}}w_{k}u\left(\sum_{j=1}^{\left|\Theta\right|}\check{q}_{j}\mathcal{I}_{k}\left(\mathcal{V}_{j}\right)\right)\right\} ,\label{eq:Pdet}\\
 & \textrm{s.t.}\:\Theta\subseteq\Theta_{T}\left(\mathcal{G}_{T}\right),\left|\Theta\right|\leq\left|\mathcal{U}_{B}\right|+1,\nonumber 
\end{eqnarray}
where $\mathcal{V}_{j}$ is the $j^{\textrm{th}}$ MIS in $\Theta$.
To complete the proof of Theorem \ref{thm:Asymptotic-equivalence-ofPB},
we only need to further prove that Algorithm B2 converges to the optimal
solution of problem (\ref{eq:Pdet}). Using the fact that any point
in a $\left(\left|\mathcal{U}_{B}\right|-1\right)$-dimensional convex
polytope can be expressed as a convex combination of no more than
$\left|\mathcal{U}_{B}\right|$ vertices, it can be shown that there
are at most $\left|\mathcal{U}_{B}\right|$ non-zero elements in $\check{\mathbf{q}}^{(i+1)}$
in step 1 of Algorithm B2. Hence $\left|\Theta^{(i)}\right|\leq\left|\mathcal{U}_{B}\right|+1,\:\forall i$.
Moreover, it can be verified that $\check{U}_{B}\left(\Theta^{(i+1)}\right)>\check{U}_{B}\left(\Theta^{(i)}\right)$
if $\Theta^{(i)}$ is not optimal for (\ref{eq:Pdet}). Combining
the above and the fact that $\check{U}_{B}\left(\Theta\right)$ is
upper bounded by $\check{U}_{B}\left(\Theta_{T}\left(\mathcal{G}_{T}\right)\right)$,
Algorithm B2 must converge to the optimal solution of (\ref{eq:Pdet}).
This completes the proof.\end{IEEEproof}
\begin{rem}
In step 2 of Algorithm B2, problem (\ref{eq:MWISP}) is equivalent
to finding a \textit{maximum weighted independent set} (MWIS) in the
interference graph $\mathcal{G}_{I}\left(\mathcal{G}_{T}\right)$
with the weights of the vertex nodes given by $\check{\mu}_{k}^{(i)}\textrm{log}\left(1+P_{n}\sigma_{k,n}^{2}\right)$.
The MWIS problem has been well studied in the literature \cite{Brendel_NIPS10_MWIS}.
Although it is in general NP hard, there exists low complexity algorithms
for finding near-optimal solutions \cite{Brendel_NIPS10_MWIS}. Although
the Asymptotic global optimality of Algorithm B2 is not guaranteed
when step 2 is replaced by a low complexity solution of (\ref{eq:MWISP}),
we can still prove its monotone convergence.
\end{rem}



\begin{thebibliography}{10}
\providecommand{\url}[1]{#1}
\csname url@samestyle\endcsname
\providecommand{\newblock}{\relax}
\providecommand{\bibinfo}[2]{#2}
\providecommand{\BIBentrySTDinterwordspacing}{\spaceskip=0pt\relax}
\providecommand{\BIBentryALTinterwordstretchfactor}{4}
\providecommand{\BIBentryALTinterwordspacing}{\spaceskip=\fontdimen2\font plus
\BIBentryALTinterwordstretchfactor\fontdimen3\font minus
  \fontdimen4\font\relax}
\providecommand{\BIBforeignlanguage}[2]{{%
\expandafter\ifx\csname l@#1\endcsname\relax
\typeout{** WARNING: IEEEtran.bst: No hyphenation pattern has been}%
\typeout{** loaded for the language `#1'. Using the pattern for}%
\typeout{** the default language instead.}%
\else
\language=\csname l@#1\endcsname
\fi
#2}}
\providecommand{\BIBdecl}{\relax}
\BIBdecl

\bibitem{Ebrahimi_TIT07_scalecell}
M.~Ebrahimi, M.~Maddah-Ali, and A.~Khandani, ``Throughput scaling laws for
  wireless networks with fading channels,'' \emph{IEEE Trans. Inf. Theory},
  vol.~53, no.~11, pp. 4250 -- 4254, Nov. 2007.

\bibitem{Gesbert_TIT11_scalecell}
D.~Gesbert and M.~Kountouris, ``Rate scaling laws in multicell networks under
  distributed power control and user scheduling,'' \emph{IEEE Trans. Inf.
  Theory}, vol.~57, no.~1, pp. 234 -- 244, Jan. 2011.

\bibitem{gesbert2007adaptation}
D.~Gesbert, S.~Kiani, A.~Gjendemsj \emph{et~al.}, ``Adaptation, coordination,
  and distributed resource allocation in interference-limited wireless
  networks,'' \emph{Proceedings of the IEEE}, vol.~95, no.~12, pp. 2393--2409,
  2007.

\bibitem{altman2006survey}
E.~Altman, T.~Boulogne, R.~El-Azouzi, T.~Jimenez, and L.~Wynter, ``A survey on
  networking games in telecommunications,'' \emph{Computers \& Operations
  Research}, vol.~33, no.~2, pp. 286--311, 2006.

\bibitem{Stolyar:2008fk}
A.~L. Stolyar and H.~Viswanathan, ``Self-organizing dynamic fractional
  frequency reuse in ofdma systems,'' in \emph{INFOCOM 2008. The 27th
  Conference on Computer Communications. IEEE}.\hskip 1em plus 0.5em minus
  0.4em\relax IEEE, 2008, pp. 691--699.

\bibitem{irmer2011coordinated}
R.~Irmer, H.~Droste, P.~Marsch, M.~Grieger, G.~Fettweis, S.~Brueck, H.~Mayer,
  L.~Thiele, and V.~Jungnickel, ``Coordinated multipoint: Concepts,
  performance, and field trial results,'' \emph{IEEE Communications Magazine},
  vol.~49, no.~2, pp. 102--111, 2011.

\bibitem{dahrouj2010coordinated}
H.~Dahrouj and W.~Yu, ``Coordinated beamforming for the multicell multi-antenna
  wireless system,'' \emph{IEEE Transactions on Wireless Communications},
  vol.~9, no.~5, pp. 1748--1759, 2010.

\bibitem{shi2011iteratively}
Q.~Shi, M.~Razaviyayn, Z.-Q. Luo, and C.~He, ``An iteratively weighted {MMSE}
  approach to distributed sum-utility maximization for a {MIMO} interfering
  broadcast channel,'' \emph{IEEE Trans. Signal Processing}, vol.~59, no.~9,
  pp. 4331--4340, sept. 2011.

\bibitem{Qualcomm2012_lte}
\emph{LTE Advanced: Heterogeneous Networks}, Qualcomm Incorporated, 2010.

\bibitem{3gpp_Rel10}
\BIBentryALTinterwordspacing
\emph{E-UTRA; Further Advancements for {E-UTRA} Physical Layer Aspects}, 3GPP
  TR 36.814. [Online]. Available: \url{http://www.3gpp.org}
\BIBentrySTDinterwordspacing

\bibitem{Wang:2012uq}
Y.~Wang and K.~I. Pedersen, ``Performance analysis of enhanced inter-cell
  interference coordination in lte-advanced heterogeneous networks,'' in
  \emph{Vehicular Technology Conference (VTC Spring), 2012 IEEE 75th}.\hskip
  1em plus 0.5em minus 0.4em\relax IEEE, 2012, pp. 1--5.

\bibitem{Pang:2012kx}
J.~Pang, J.~Wang, D.~Wang, G.~Shen, Q.~Jiang, and J.~Liu, ``Optimized
  time-domain resource partitioning for enhanced inter-cell interference
  coordination in heterogeneous networks,'' in \emph{Wireless Communications
  and Networking Conference (WCNC), 2012 IEEE}.\hskip 1em plus 0.5em minus
  0.4em\relax IEEE, 2012, pp. 1613--1617.

\bibitem{Luo_Arxiv12_WMMSE_HetNet}
\BIBentryALTinterwordspacing
M.~Hong, R.-Y. Sun, H.~Baligh, and Z.-Q. Luo, ``Joint base station clustering
  and beamformer design for partial coordinated transmission in heterogenous
  networks,'' 2012. [Online]. Available: \url{http://arxiv.org/abs/1203.6390}
\BIBentrySTDinterwordspacing

\bibitem{Mo_TACM00_alfafair}
J.~Mo and J.~Walrand, ``Fair end-to-end window-based congestion control,''
  \emph{IEEE/ACM Transactions on Networking}, vol.~8, no.~5, pp. 556--567, Oct
  2000.

\bibitem{Kelly_OPR98_PFS}
F.~Kelly, A.~Maulloo, and D.~Tan, ``Rate control for communication networks:
  Shadow price proportional fairness and stability,'' \emph{J. Oper. Res.
  Soc.}, vol.~49, pp. 237--252, 1998.

\bibitem{Whiting_TWC04_PFSconv}
H.~Kushner and P.~Whiting, ``Convergence of proportional-fair sharing
  algorithms under general conditions,'' \emph{IEEE Transactions on Wireless
  Communications}, vol.~3, no.~4, pp. 1250--1259, 2004.

\bibitem{ghaffar2010fractional}
R.~Ghaffar and R.~Knopp, ``Fractional frequency reuse and interference
  suppression for {OFDMA} networks,'' in \emph{Proceedings of the 8th
  International Symposium on Modeling and Optimization in Mobile, Ad Hoc and
  Wireless Networks}, 2010, pp. 273--277.

\bibitem{somekh2009cooperative}
O.~Somekh, O.~Simeone, Y.~Bar-Ness, A.~Haimovich, and S.~Shamai, ``Cooperative
  multicell zero-forcing beamforming in cellular downlink channels,''
  \emph{IEEE Trans. Inf. Theory}, vol.~55, no.~7, pp. 3206--3219, 2009.

\bibitem{Brendel_NIPS10_MWIS}
W.~Brendel and S.~Todorovic, ``Segmentation as maximum weight independent
  set,'' \emph{in NIPS}, pp. 307--315, 2010.

\bibitem{Boyd_04Book_Convex_optimization}
S.~Boyd and L.~Vandenberghe, \emph{Convex Optimization}.\hskip 1em plus 0.5em
  minus 0.4em\relax Cambridge University Press, 2004.

\end{thebibliography}
\end{document}